\documentclass[a4paper,fleqn,usenatbib]{mnras}

\usepackage[T1]{fontenc}
\usepackage{ae,aecompl}
\usepackage{graphicx}
\usepackage{amsmath}
\usepackage{amssymb}
\usepackage{tablefootnote}
\usepackage{threeparttable}

\title[Mass function of open clusters]{The family pictures of our neighbours: investigating the mass function and dynamical parameters of nearby open clusters}

\author[Ebrahimi et al.]
{H. Ebrahimi$^{1}$\thanks{E-mail:  \mbox{h.ebrahimi@iasbs.ac.ir} (HE)},
A. Sollima$^{2}$,  
H. Haghi$^{1}$ \\
$^{1}$Department of Physics, Institute for Advanced Studies in Basic Sciences (IASBS), Zanjan 45137-66731, Iran\\
$^{2}$INAF Osservatorio Astrofisica e Scienza dello Spazio, via Gobetti 93/3, I-Bologna 40129, Italy }
\begin{document}

\date{Accepted .... Received}

\pagerange{\pageref{firstpage}--\pageref{lastpage}} \pubyear{2022}

\maketitle

\label{firstpage}

\maketitle

\begin{abstract}

We determine the mass functions (MFs) and the dynamical parameters of 15 nearby open clusters (OCs) using the unprecedented data set of the Gaia Early Data Release 3. 
We select the members of each cluster by combining the photometric (colour and magnitude) and astrometric (parallax and proper motions) parameters of stars, minimizing the contamination from Galactic field interlopers. 
By comparing the observed distribution of stars along the cluster main sequence with the best-fitting synthetic population, we find the present-day MF and the binary fraction of the OCs, along with their dynamical parameters like mass, half-mass radius, and half-mass relaxation time. 
We found that the global present-day MF of OCs are consistent with a single power-law function, $F(m)\propto m^\alpha$, with slopes $-3<\alpha<-0.6$ including both subsolar, $0.2<m/\text{M}_\odot<1$, and supersolar mass regimes. 
A significant correlation between the MF-slope and the ratio of age to half-mass relaxation time is evidenced, similarly to the same conclusion already observed among Galactic globular clusters. 
However, OCs evolve along different tracks in comparison with the globular clusters, possibly indicating primordial differences in their initial mass function (IMF). 
The comparison with Monte Carlo simulations suggests that all the analysed OCs could have been born with an IMF with slope $\alpha_{\text{IMF}}<-2.3$. 
We also show that the less evolved OCs have a MF consistent with that of the solar neighbourhood, indicating a possible connection between the dissolution of OCs and the formation of the Galactic disc.

\end{abstract}


\section{Introduction} \label{Sec:1}

OCs and associations are some of the most interesting objects which populate our neighbours.  
They are key constituents in the study of many fields of astrophysics, from star formation and stellar evolutionary models to galactic  evolution. In this regard, the Galactic OCs are used to calibrate stellar evolutionary models at young and intermediate ages (<2 Gyr). They are among the best tools to probe the structure of the disk of our Galaxy.

More than two thousand catalogued OCs are distributed within 2 kpc to the Sun \citep{cantat-gaudin2018,zhong2020}. 
They are excellent tracers of the spiral-arms structure \citep{dias2005} and of the chemical evolution of the disc \citep{magrini2017}. 
Virtually all stars form in embedded clusters \citep{Kroupa1995a, lada2003}. These expel most of the unused gas and then lose more than 50\% of their stars \citep{Brinkmann2017}. Thus, most stars in the field come from this process. The surviving OCs slowly dissolve through the energy equipartition process releasing their stars into the Galactic field through thin tidal tails.
Therefore, the study of the present-day stellar content of the OCs helps us to understand the evolution of the Galactic disc and the solar neighbourhood. In comparison with older objects like globular clusters (GCs), OCs are relatively young \citep[$6<\log(t_{\text{age}}/\text{yr})<10$;][]{piskunov2018} and metal-rich objects \citep[$-0.5\lesssim \text{[Fe/H]}/\text{dex}\lesssim 0.5$;][]{magrini2017}. 
A typical OC contains $\approx 10^2$ to $10^4$ gravitationally bound stars 
confined within a limiting radius of $\approx 10 \, \text{pc}$ \citep{binney2008}. 
So, the typical relaxation time (when collisions become effective in exchanging energies among cluster members) of OCs is of the order of $\approx 10^7$ to $10^8$ yr, i.e. shorter than or comparable with their ages.
The structure and internal kinematics of OCs can be therefore used as a 
test for the stellar and dynamical evolution models of collisional systems.

An important information that can be extracted from the OCs colour-magnitude diagram (CMD) 
is the present-day mass function (PDMF), defined as the relative proportion of cluster stars with different masses. 
This quantity indeed reflects the initial physical conditions (density and metal content) affecting the 
efficiency of fragmentation of the primordial cloud from which these clusters form \citep{hennebelle2008}.
OCs and associations are excellent sites to measure the PDMF. They are indeed composed by coeval and chemically homogeneous 
stars located at the same distance, thus eliminating the uncertainties linked to the relative distance and to the differences in the age and chemical composition of individual stars.
In young associations, the PDMF is similar to its initial form at birth time, so called {\it initial mass function} (IMF).
In the hypothesis of an important contribution of young clusters in the formation of the Galactic disc, the 
measure of their PDMF provides crucial insight into the determination of the IMF of the disc. 

Instead, because of the effect of stellar and dynamical evolution, the PDMF of OCs is generally different from the IMF. 
The mass distribution of the main sequence (MS) stars for the OCs covers a wide range from the mass corresponding to the hydrogen 
burning limit ($\approx 0.1 \, \text{M}_\odot$) to the mass of the currently evolving stars.
As time proceeds, this upper limit decreases thus depleting the PDMF at its upper boundary.
Therefore, the high-mass end of the PDMF differs from cluster to cluster 
depending on the age and metallicity of each OC \citep{gaia2018}. 
Moreover, in the collisional multimass systems like OCs, the tendency toward the kinetic energy equipartition leads to a process,
so-called {\it dynamical mass segregation}, in which the high-mass stars sink to the central part of the cluster and low-mass stars
drift to the outskirts \citep{spitzer1987}.

Once an OC revirialises after gas expulsion, it quickly mass segregates (if it was not already mass segregated at its birth) and will consequently lose preferentially low-mass stars and depletes the low-mass end of the mass function \citep{kruijssen2009, Zonoozi2011, Zonoozi2014, Zonoozi2017, Haghi2015}. For the above reasons, in stellar systems where two-body relaxation has been effective, the PDMF appears depleted with respect to the IMF and varies across the cluster extent. The 700 Myr old Hyades cluster, for example, is significantly depleted at the low-mass end \citep{Kroupa1995b}, while the 100 Myr old Pleiades is not \citep{KroupaAH2001}.

One of the convenient ways for converting the observational luminosity/absolute magnitude distribution
of the MS stars into masses is its comparison with theoretical isochrones \citep{bergbusch1992}.
Inside a cluster, stars can be indeed considered part of a single stellar population.
At a finer sight, a few levels of complexity need to be considered to make an accurate estimate of the PDMF.
One is given by the presence of binaries, whose luminosity equals the sum of the luminosity of their components and therefore
appears brighter than single stars on the CMD.
Assuming wrongfully a massive star which could be originally the combination of the two
less massive components, causes the incorrect flattening of the PDMF \citep{kroupa1993}.
While this effect is small in GCs, it cannot be neglected in OCs where the binary fraction is generally high, e.g. $\approx 50\%$ \citep{sollima2010,sharma2008}. 
To avoid this bias, more sophisticated techniques of population synthesis are required.
Another difficulty is linked to the strong field contamination. Indeed, because of their proximity and intrinsic size, OCs extends 
over wide areas on the sky with a projected density of stars similar to the Galactic field well within their tidal radius.
Moreover, being formed by the same stellar population of the disc, 
their CMDs are quite similar to that of the surrounding field.
So, the CMD of OCs needs to be cleaned either using additional information (like proper motions and/or parallaxes) or
statistically (correcting star counts for an average background observed in a control field). 

In recent decades, many works have estimated the PDMF of OCs and compared them with the IMF of the solar neighbourhood. 
Some of them found that the PDMF of the OCs are shallower than the mentioned IMF in the subsolar regime, particularly
for the old ($t_{\text{age}}>300 \, \text{Myr}$) OCs, e.g. Coma Berenices \citep{kraus2007} and 
Hyades \citep{goldman2013}. In the supersolar regime, the PDMF of the majority of OCs is instead similar to the IMF of the solar neighbourhood,
in e.g. the Hyades \citep{bouvier2008}, the Pleiades \citep{roser2020}, and in 12 OCs located in Sagittarius 
spiral arm \citep{angelo2019}. The flattened shape of the PDMF can be interpreted as the effect of dynamical evolution.

The clear evidence of dynamical mass segregation is widely discussed 
by many papers, especially for older OCs, e.g. Praesepe \citep{khalaj2013}, Hyades \citep{roser2011}, and
Coma Berenices \citep{tang2018}. Moreover, some studies have confirmed that the mass segregation is even
visible in younger OCs with $t_\text{age}<50 \, \text{Myr}$ \citep{schilbach2006} and interpreted it as an evidence of
primordial mass segregation within the protostellar gas cloud \citep{maciejewski2007}. 

Several studies investigated the relation between the PDMF slope and the relaxation process 
by comparing the observational data with dynamical models in collisional stellar systems like GCs and OCs.
Among GCs, a tight correlation between the PDMF slope with 
the ratio of the age to the relaxation time has been found, suggesting that the internal dynamical evolution
governs the shape of PDMF \citep{sollima2017, baumgardt2018, ebrahimi2020}. 
A similar discussion for younger
objects like Galactic OCs is a matter of debate in recent years. \cite{bonatto2005} derived the global PDMF
of 11 nearby OCs and fitted them with a broken power-law function with different indices at low- and high-mass regimes. They
showed that both PDMF parts flatten in OCs with age older than the relaxation time. \cite{maciejewski2007}
expanded the sample to 42 Galactic OCs and showed that no correlation exists between global PDMF and
the ratio of age to relaxation time. The flattening of PDMF is only clearly visible in their sample for the OCs with ages
older than $\approx$100 times the relaxation time. Moreover, \cite{sharma2008} found that the
global PDMF of nine OCs flatten with increasing the ratio of age to relaxation time.

The unprecedented data set of more than one billion sources which have been published through the third               
\emph{Gaia} data release (EDR3), pushed the boundaries of the new astronomy. The \emph{Gaia} EDR3 contains the
precise photometric parameters, e.g. colours and magnitudes, which are complete averagely over the $G<20$ magnitude across the entire sky. 
In addition, it includes the astrometric parameters like positions, proper motions, and parallaxes with
an average systematic uncertainty < 0.05 mas even for relatively faint stars \citep[$G<20$;][]{gaia2021}. The quality of the
\emph{Gaia} data enable the studies to discover new OCs \citep{castro-ginard2020}, reclassify some of them \citep{cantat-gaudin2020}, 
and improve the properties of the catalogued OCs \citep{bossini2019}, allowing to use of proper motions and parallaxes as excellent tools to separate members from field interlopers \citep{gaia2018}.
With the help of the Gaia mission,  the discovery of hundreds of new clusters and the characterisation of extended stellar structures identified either as relics of filamentary star formation or tidal tails have been achieved. Moreover,  the detailed study of their properties, such as age, chemical composition and orbits have been investigated.

In this paper, we use the \emph{Gaia} EDR3 dataset to derive the PDMF of 15 nearby OCs through a population synthesis approach. 
In Section 2, we describe the selection
criteria to determine the membership of OCs from the full catalogue of \emph{Gaia} data and define a recipe
to take into account the effect of contamination. In Section 3, we explain the algorithm to estimate the PDMF of OCs using synthetic population. The PDMF of our OCs sample
and the possible correlations with the various parameters are presented in Section 4.
Finally, we discuss our results in Section 5.

\section{Observational Data Analysis} \label{Sec:2} 

\subsection{Data and Targets} \label{Sec:2.1}

In the present analysis, the third Early-Data Release (EDR3) of \emph{Gaia} has been used. 
For each OC, We retrieved from the Gaia archive\footnote{https://gea.esac.esa.int/archive/}\citep{gaia2021} 
magnitudes in $G$, $G_{BP}$, and $G_{RP}$ passbands, 
positions ($RA$ and $Dec$), proper motions ($\mu^*_{RA}$, $\mu_{Dec}$), and parallaxes ($p$), and their related uncertainties. 
The catalogues
including the photometric and astrometric solutions have been retrieved within circular fields around the centre of each OC.
The radius of the circular field has been chosen to cover the entire cluster population and to sample that of the surrounding Galactic field.
It ranges between 15 and 30 deg which is $\approx$1.5--3 times the tidal radius of each OC (5--20 deg).

We selected 15 well-defined target OCs which are located in the solar neighbourhood
between $\approx 45\, \text{pc}$ (Hyades) and $\approx415\, \text{pc}$ (NGC 2516) from the Sun. They have been chosen on the basis of both their distance and their masses to
ensure a good statistical sampling across the entire stellar mass range, including the faint/low-mass stars close to the hydrogen-burning limit. 
The OCs analysed in the present work are Hyades, Coma Berenices (Melotte 111), Pleiades (Messier 45, Melotte 22),
IC 2391, IC 2602, Alpha Persei/$\alpha \, \text{Per}$ (Melotte 20), Praesepe (NGC 2632, Messier 44), NGC 2451A,
Blanco 1, NGC 6475 (Messier 7), NGC 7092 (Messier 39), NGC 6774 (Ruprecht 147), NGC 2232, NGC 2547, NGC 2516.  
Our selected target OCs all have well populated CMDs and are not influenced by significant interstellar reddening variations \citep{gaia2018}.
Our OCs sample covers a wide range of ages $\approx 40 \, \text{Myr}-2.65 \, \text{Gyr}$ \citep{gaia2018, pang2021} 
which enables us to comprehensively cover both the short and long-term evolution of OCs. The age, metallicity, distance modulus , and reddening of OCs in our sample which are applied in the membership determination process and synthetic stars construction are listed in Table \ref{table_structure}.

\begin{table}
    \centering
    \caption{Age (2nd column), metallicity (3rd column), distance modulus (4th column), and reddening (5th column) of OCs in our sample.}
    \label{table_structure}
    \begin{tabular}{cccccccccc}  
    \hline     
    Cluster & $\log(\text{age/yr})$ & [Fe/H]  & DM & $E(B-V)$ \\
    \hline
    Hyades         & 8.90  & 0.13  & 3.389  & 0.001 \\
    Coma Berenices & 8.81  & 0.00  & 4.669  & 0.000 \\
    Pleiades       & 8.04  & -0.01 & 5.667  & 0.045 \\
    IC 2391        & 7.70  & -0.01 & 5.908  & 0.030 \\
    IC 2602        & 7.60  & -0.02 & 5.914  & 0.031 \\
    $\alpha$-Per   & 7.85  & 0.14  & 6.214  & 0.090 \\
    Praesepe       & 8.85  & 0.16  & 6.350  & 0.027 \\
    NGC 2451A      & 7.78  & -0.08 & 6.433  & 0.000 \\
    Blanco 1       & 8.06  & 0.03  & 6.876  & 0.010 \\
    NGC 6475       & 8.54  & 0.02  & 7.234  & 0.049 \\
    NGC 7092       & 8.54  & 0.00  & 7.390  & 0.010 \\
    NGC 6774       & 9.42* & 0.16  & 7.455  & 0.110*\\
    NGC 2232       & 7.70  & 0.11  & 7.575  & 0.031 \\
    NGC 2547       & 7.60  & -0.14 & 7.980  & 0.040 \\
    NGC 2516       & 8.48  & 0.05  & 8.091  & 0.071 \\
    \hline  
    \end{tabular} 
     \begin{tablenotes}
      \item \textbf{Notes.} The age and distance modulus of NGC 6774 marked by (*) are adopted from \cite{pang2021}. The rest of values are derived from \cite{gaia2018}.   
      \end{tablenotes}
\end{table}

\begin{figure*}
\begin{center}
\includegraphics[width=180mm]{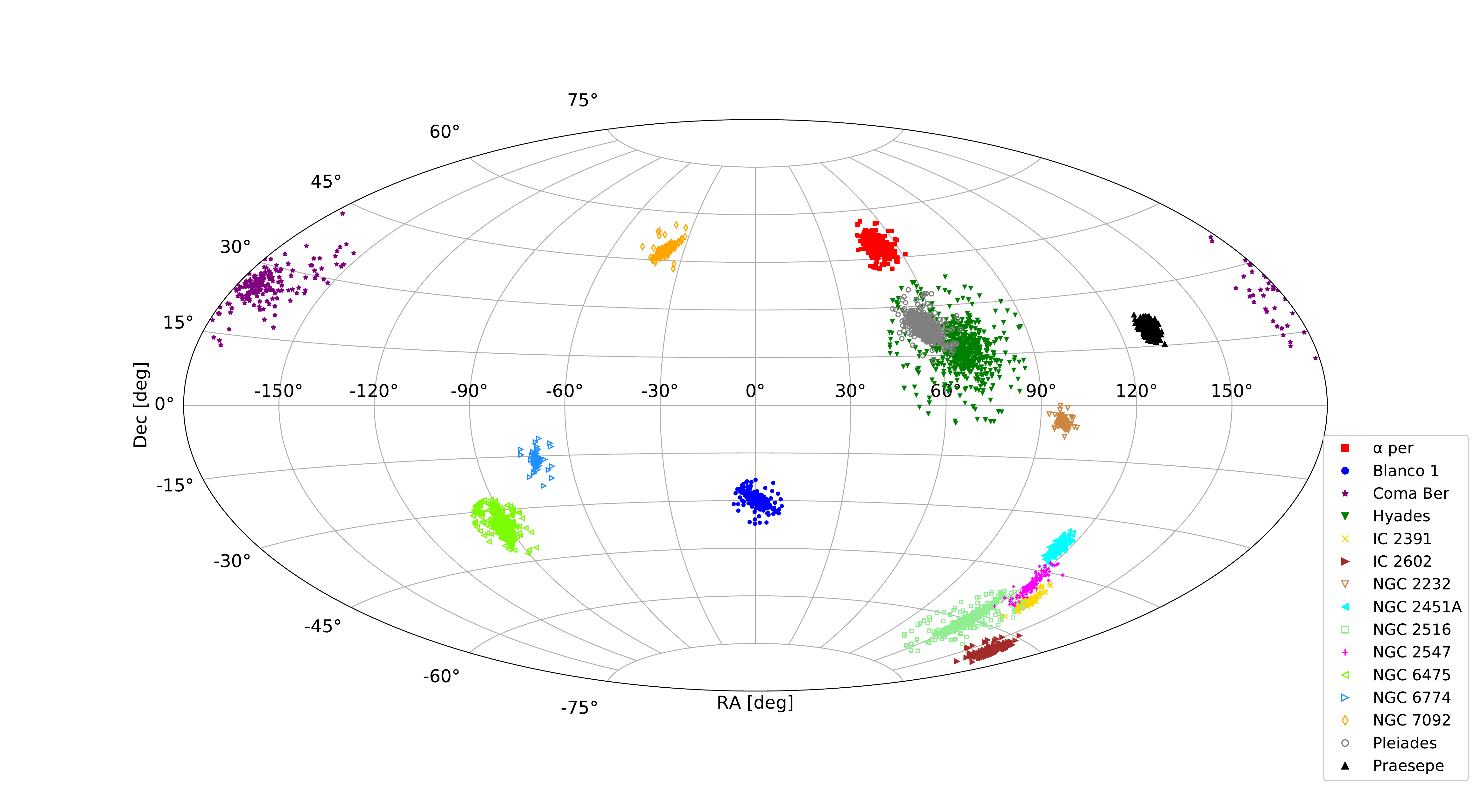}
\caption{Spatial position distributions of the bona-fide members of all OCs sample in celestial coordinate $(RA, Dec)$. The members of the 15 OCs inside the tidal are shown with different colours and symbols.}
\label{map}
\end{center}
\end{figure*}

\begin{figure*}
\begin{center}
\includegraphics[width=150mm]{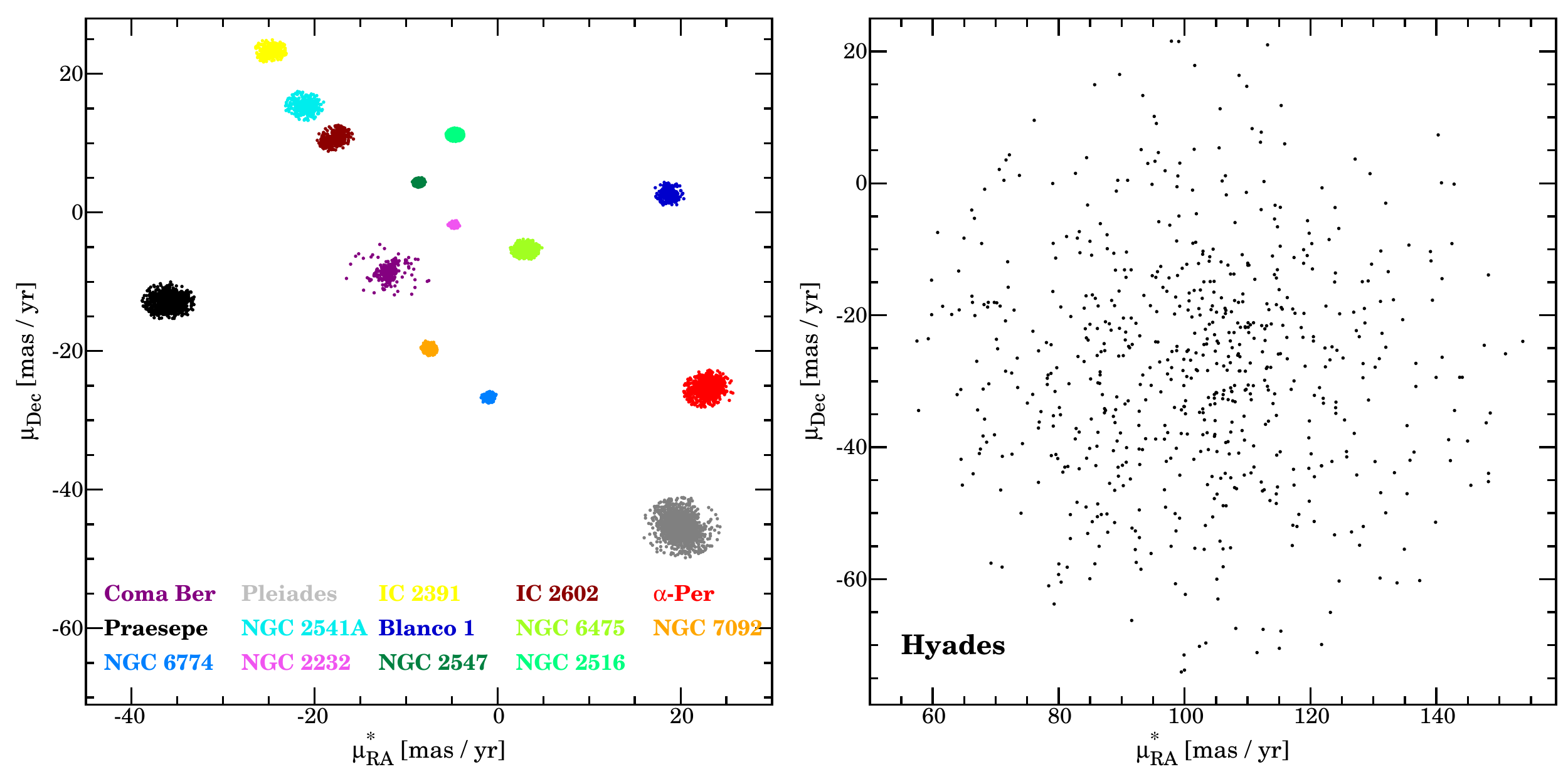}
\caption{Distribution of 14 OCs target (left panel) and Hyades (right panel) in the proper motions plane. The members inside the tidal radius for each OC are shown in different colours.}
\label{pm}
\end{center}
\end{figure*} 

\begin{figure*}
\begin{center}
\includegraphics[width=170mm]{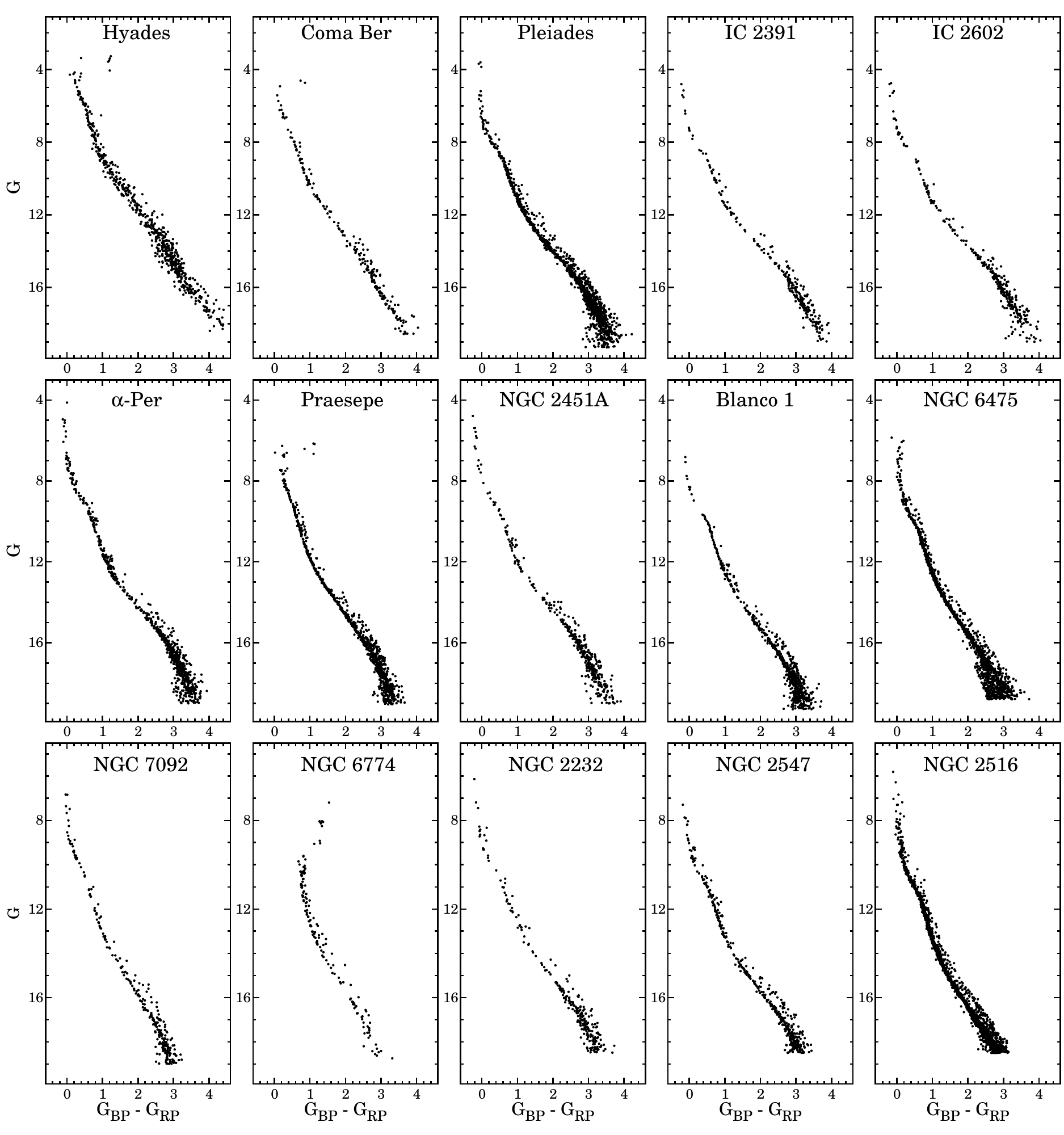}
\caption{CMDs of the 15 OCs of our sample. The members inside the tidal radius for each OC are shown.}
\label{cmd}
\end{center}
\end{figure*}
\subsection{Membership Determination} \label{Sec:2.2}
The most difficult task in the determination of the MF in OCs is linked to the contamination from the Galactic field interlopers.
The procedure for membership identification has been performed in three steps.

In the first step, we define a mean ridge line in the CMD and determine the mean proper motions and parallax of the target OC.
In the first iteration, we use all the stars located within 2 deg from the cluster centre, while in the subsequent ones only bona-fide cluster 
members inside the tidal radius have been used. 

In the second step, we select first-guess members selecting stars in the 5D parameter space ($G$, $G_{BP}-G_{BP}$, $\mu^*_{RA}$, $\mu_{Dec}$, $p$) 
within a given distance from the CMD mean ridge line and within a given distance from the bulk proper motion and parallax. The $G$ magnitude-dependent boundaries of the selection box in $G_{BP}-G_{RP}$ colour ($\sigma_{col}$), proper motion ($\sigma_{\mu}$) and parallax ($\sigma_p$) have been determined using 
the post-release error formulas proposed exclusively for EDR3 of the \emph{Gaia} mission\footnote{The corresponding error formulas for 
magnitudes, colour, proper motions and parallaxes have been listed in the following website: https://www.cosmos.esa.int/web/gaia/science-performance.}. 
Whenever the Johnson-Cousins related colour ($V-I_C$) appears in these formulas, we interpolate it through the polynomial colour-colour 
transformation to the \emph{Gaia}-based colour ($G_{BP}-G_{RP}$) presented by \cite{jordi2010}. As these formulas often provide a non-optimal description of the actual observational error, 
we correlate them with the formal errors reported in the Gaia catalogue and construct linear fits for all the five considered parameters. 
These relations have been used to correct the boundaries of the selection boxes. 
For the CMD, the red side of the selection box has been shifted by 0.752 mag toward brighter magnitudes, to include binaries (see Section \ref{Sec:2.3} for details).  
A lower limit has been also imposed at $G<20$, to limit the analysis only to stars in the $G$ magnitude range with high completeness. 
For the proper motion, because of the motion of the stars inside the cluster, there is also an intrinsic spread in velocity
which must be summed in quadrature to all member stars. To include this effect, we calculate the standard deviation of
both components of proper motion and parallax ($\sigma_{\mu,\text{min}},~\sigma_{p,\text{min}}$) for very bright stars ($G<10$), for which
the observational errors are negligible. Thus, the total uncertainties of proper motions
can be written as $\sigma_{\mu}=\sqrt{\sigma^2_{\mu,0}+\sigma^2_{\mu,\text{min}}}$ and $\sigma_p=\sqrt{\sigma^2_{p,0}+\sigma^2_{p,\text{min}}}$, where $\sigma_{\mu,0}$ and $\sigma_{p,0}$ are the uncertainties in proper motions and parallax estimated using the Gaia error formulas. 
We finally select all stars lying within $n$ times the magnitude-dependent 
error in all parameters. We tune the free parameters set ($n_{col},~n_{\mu},~n_p$) to include the bulk of the stars 
observed inside the tidal radius. 
The density of stars passing all the above criteria located outside the tidal radius has been estimated and adopted as the density of residual contamination.

As a third step, the observed cumulative distribution of distances of member stars from the centre of the OC has been compared 
with those predicted by a set of theoretical \cite{king1966} models (after accounting for the constant density of field contaminants estimated above) and the values of the central adimensional potential ($W_{0}$) and core radius ($r_{c}$) 
providing the minimum difference between model and observation have been determined. 
We estimate the tidal radius of each OC using the corresponding best-fit King model parameters.

The above three steps have been repeated iteratively, re-calculating the mean proper motion and parallax using only the members selected in the previous iteration within the updated tidal radius (see above). The fraction of selected members within the tidal radius, once correcting for residual contamination, ranges from 0.001 to 0.17 per cent, indicating a strong dominance of field interlopers.

The two-dimensional spatial distributions of the selceted members of all the 15 target OCs are displayed in Fig. \ref{map}. 
Moreover, the distribution of the OCs in two dimensional proper motions plane are shown in Fig \ref{pm}.
The CMD of the member stars inside the tidal radius for each OC is shown in Fig. \ref{cmd}. 
  
\begin{figure}
\begin{center}
\includegraphics[width=80mm]{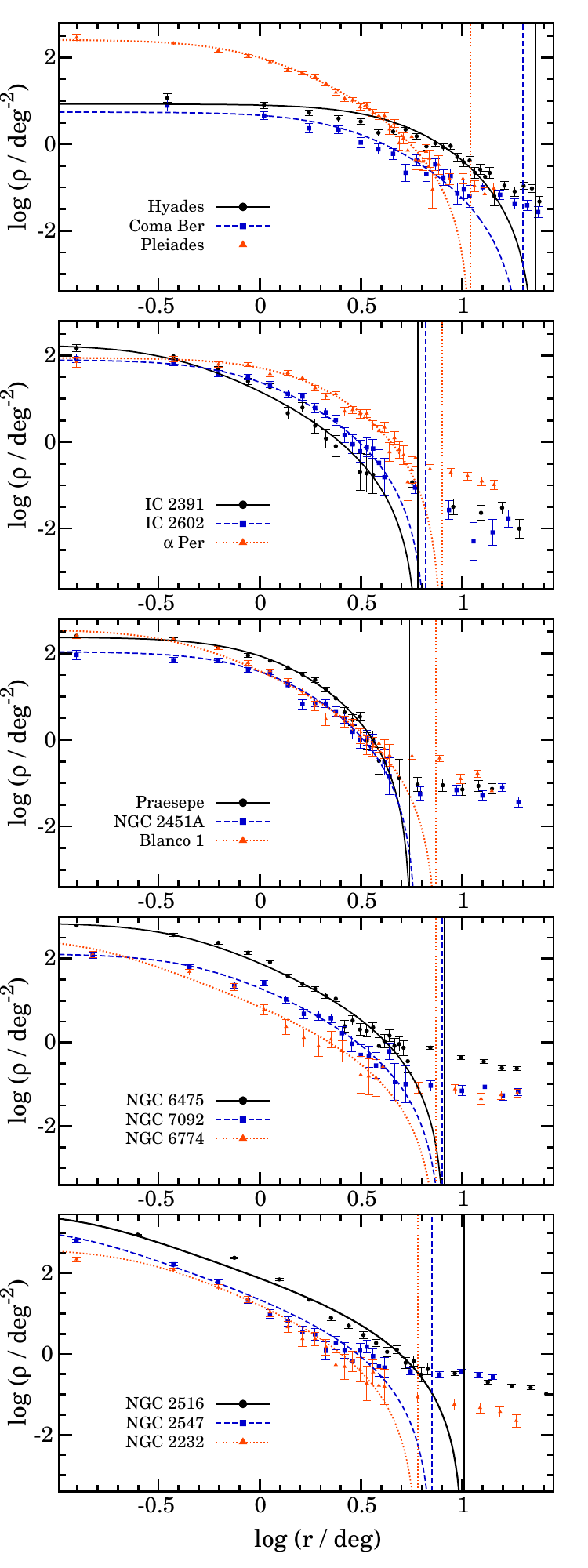}
\caption{Projected density profiles of the 15 OCs of our sample. Filled circles, squares and triangles represent the observational density profiles. The density profiles of the corresponding best-fitting models are marked with solid, dotted, and dashed lines. The vertical solid, dotted, and dashed lines mark the tidal radii.}
\label{dR}
\end{center}
\end{figure}

In Fig. \ref{dR}, we compare the observational and best-fitting density profiles of the whole OCs sample, for visualization purpose. 
The predicted projected density of the best-fitting models is generally in good agreement with the observed OCs density inside the tidal 
radii: The reduced $\chi^2$ of the fit ranges between 0.43 and 3.92 for 14 clusters of our sample except for NGC 2516 with $\chi^2=14.9$. In this last cluster a significant deviation from the sudden decline predicted by the best fit \cite{king1966} model is apparent, possibly indicating a significant tidal heating in the outskirts of this cluster.
Outside the tidal radius, the density continues to decline and flattens as a result of the joint effect of extra-tidal stars and 
field interlopers. In older OCs, i.e. Coma Berenices and Hyades, the density level at the core is around 100 times higher than that at the coronae 
while it reaches even larger values of $10^3-10^4$ for the rest of the sample. 
The tidal radii of our OCs target range from $r_t=5.5 \, \text{deg}$ (Praesepe) to $r_t=22.9 \, \text{deg}$ (Hyades). 

\begin{figure*}
\begin{center}
\includegraphics[width=155mm]{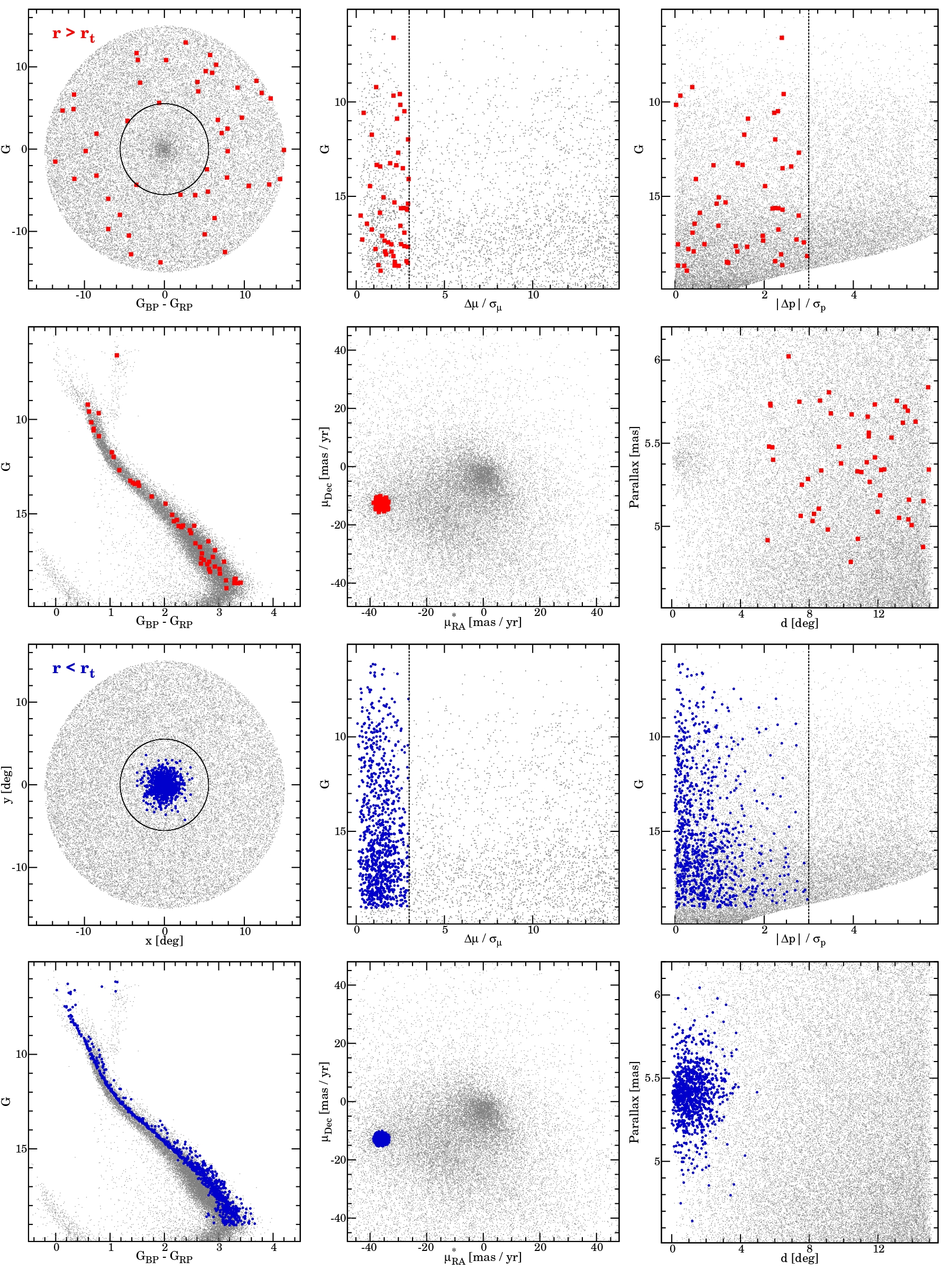}
\caption{Distribution of all \emph{Gaia} sources with parallaxes $4.5 < p/\text{mas} < 6.2$ and distances from the Praesepe centre $d<15 \, \text{deg}$ in the colour-magnitude (Left panels in the first and third rows), proper motions (middle panels in the first and third rows), $d$-parallax (right panels in the first and third rows), projected positions or $x-y$ (left panels in the second and fourth rows), proper motion selection box (middle panels in the second and fourth rows), and parallax selection box (right panels in the second and fourth rows) diagrams. The grey dots show the background sources and blue circles and red squares represent the selected members inside and outside the tidal radius, respectively, for Praesepe. The dashed lines in the selection box plots represent the boundaries of the selection criteria and black circles in position planes identify the tidal radii.}
\label{criteria}
\end{center}
\end{figure*} 

In Fig \ref{criteria}, the selected bona-fide members of Praesepe inside and outside the tidal radius are shown, as an example.
It can be noted that, while the adopted selection criteria properly identify the majority of Galactic field interlopers, still a few stars 
evenly distributed around the cluster well above its tidal radius pass the selection criteria.
These objects are field stars with a distribution in parallax, proper motion and CMD overlapping with those of cluster stars or, in a small fraction, to members recently escaped from the cluster.
This indicates that the adopted criteria, while valuable, admit the residual contamination from field contaminants.
We used the MF of stars outside the tidal radius spuriously classified as members to decontaminate the cluster MF (see Sect. \ref{Sec:2.3}).    

\subsection{Binary Fraction and Mass Function} \label{Sec:2.3}

The high quality of the \emph{Gaia} photometric data can considerably disclose the location of the binaries on the OCs CMD. 
The flux of a binary system is measurable as the sum of the fluxes of its components. 
Thus, the binaries intrinsically appear brighter than the MS single stars on the CMD and cause shifts in their colours depending on their 
component magnitudes in each passband. If we assume a binary system including two equal-mass components, this shift will be $-2.5 \log(2)\approx 0.752 \, \text{mag}$ brighter than the MS single stars \citep{sollima2007}. 
Thus, a reference border that defines the binary box has been adopted by shifting the red boundary of the CMD selection box by $\approx 0.752$ mag brighter than the mean ridge line. 
We define as bona-fide binaries all member stars located on the blue side of this boundary and redder than the MS selection box (i.e. with $G_{BP}-G_{RP}>(G_{BP}-G_{RP})_{\text{MS}}+n_{col}\sigma_{col}$; see Fig. 5 of \cite{sollima2007}). 
Note that the binaries contained in this box are only part of the total population of binaries. Indeed, binaries with low mass ratios can 
also be located in redder colours, confusing with single stars. 

To evaluate the raw MF, we first define the nine evenly spaced mass groups in the $0.1<m/\text{M}_{\odot}<1$ range and $3-5$ mass bins in 
the $m/\text{M}_{\odot}>1$ range depending on the maximum mass of member stars in each OC. To decontaminate the MF, we correct the total number of stars counted inside each mass bin ($N_{\text{i}}$) using the star counts inside ($N_{i,\text{in}}$) and outside ($N_{i,\text{out}}$) the tidal radius as:
\begin{equation} \label{ntot}
N_{\text{i}}=N_{\text{i,in}}-N_{\text{i,out}}\bigg(\dfrac{A_{\text{in}}}{A_{\text{out}}}\bigg),
\end{equation}     
where $A_{\text{in}}=\pi r_t^2$ and $A_{\text{out}}=\pi (r_{\text{max}}^2-r_t^2)$ are the areas occupied by the members stars inside and outside the tidal radius, respectively. 
Note that, $r_{\text{max}}$ is the maximum radius of circular field of view for each OC which is retrieved from the \emph{Gaia} archive.     

To take into account the effect of contamination in estimating the binary fraction, we define the corrected total number of the star members inside the binary box as:  
\begin{equation} \label{ntot_bin}
N_{\text{bin}}=N^{\text{bin}}_{\text{in}}-N^{\text{bin}}_{\text{out}}\bigg(\dfrac{A_{\text{in}}}{A_{\text{out}}}\bigg),
\end{equation}
where $N^{\text{bin}}_{\text{in}}$ and $N^{\text{bin}}_{\text{out}}$ are the stars occupying the binary box inside and outside the tidal radius, respectively. The observational binary fraction has been then estimated as $f_b=N_{\text{bin}}/N_{\text{tot}}$, where $N_{\text{tot}}=\sum_{i} N_i$ has been determined from equation \ref{ntot}.

We construct the MF by counting the number of member stars in each mass bin and in the binary box, and correcting them using equations \ref{ntot} and \ref{ntot_bin}.

\section{Method} \label{Sec:3}

In this section, we describe the adopted iterative algorithm to estimate the global OCs MF, binary fraction, and dynamical parameters. 
This method is based on exploring the best-fit distribution of a synthetic stellar population with the observational distribution of stars on the CMD of each OC \citep[see][]{sollima2012,sollima2019}.

In the adopted algorithm, the following steps have been sequentially performed:
\begin{enumerate}
\item{ The masses of $10^5$ synthetic particles are extracted from a guess MF with masses between $0.1 \, \text{M}_{\odot}$ and the maximum mass of the observational sample.
At the first iteration, the coefficients defining the MF ($k_{i}$) follow a \citet{kroupa2001} IMF: $k_i\propto m^{-1.3}_i$ for $m_i<0.5 \, \text{M}_{\odot}$ and 
$k_i\propto m^{-2.3}_i$ for $m_i>0.5 \, \text{M}_{\odot}$.}
\item{The corresponding $G$, $G_{BP}$, and $G_{RP}$ magnitudes of the synthetic stellar population have been calculated by interpolating the 
masses through the PARSEC isochrone \cite{bressan2012}\footnote{http://stev.oapd.inaf.it/cgi-bin/cmd.} with the best-fitting metallicity, 
age, distance modulus, and reddening provided by \cite{gaia2018} and \cite{pang2021} listed in Table \ref{table_structure}.}
\item{We assumed an initial global binary fraction of 10\% as a first guess and paired a corresponding fraction of synthetic stars with secondary components extracted from the same MF. 
The total magnitude of each synthetic binary system is given by:

\begin{equation}
G^{\text{bin}}_{\text{tot}}=-2.5\log\bigg(10^{-0.4{G^{{\text{bin}}}_1}}+10^{-0.4G^{{{\text{bin}}}}_2}\bigg),
\end{equation}
where $G_1^{\text{bin}}$ and $G_2^{\text{bin}}$ are the magnitudes of the two components of a binary.

The rest of the synthetic stars are assumed to belong to the single star population. 
Similar procedures have been repeated for the $G_{BP}$ and $G_{RP}$ of the binaries to construct a complete set of the synthetic single star and binary populations with given masses, magnitudes, and colours.}

\item{For each synthetic star, the errors of $G$, $G_{BP}$, and $G_{RP}$ magnitudes have been defined using the synthetic uncertainties estimated in Sect. \ref{Sec:2.2}, according to their magnitudes. 
The $G_{BP}$ and $G_{RP}$ errors have been multiplied by a fudge factor tuned to reproduce the colour spread of the bright portion of the MS (at $10<G<15$) 
in the first iteration.}
\item{To mimic the photometric errors in $G$, $G_{BP}$, and $G_{RP}$ magnitudes, Gaussian distributed random numbers have been extracted, 
multiplied by the error of each magnitude estimated in the previous step, and added to the corresponding magnitude.}
\item{The effect of incompleteness has been simulated using the \textsc{selectionfunctions} package presented by \cite{boubert2020} for \emph{Gaia} EDR3\footnote{A corresponding Python package \textsc{selectionfunctions} is available in the following website: https://github.com/
gaiaverse/selectionfunctions.}. Using the above package, we estimate the completeness of each synthetic particle, and retain the star if a random number uniformly 
distributed between 0 and 1 was smaller than the corresponding completeness.}
\item{We select the suitable synthetic stars by employing the same CMD selection box adopted for observations (see Section \ref{Sec:2.2}). 
The synthetic raw MF has been derived by counting single and binary synthetic stars in the same bins defined in Section \ref{Sec:2.3} for observations.} 
\end{enumerate} 

\begin{figure*}
\begin{center}
\includegraphics[width=130mm]{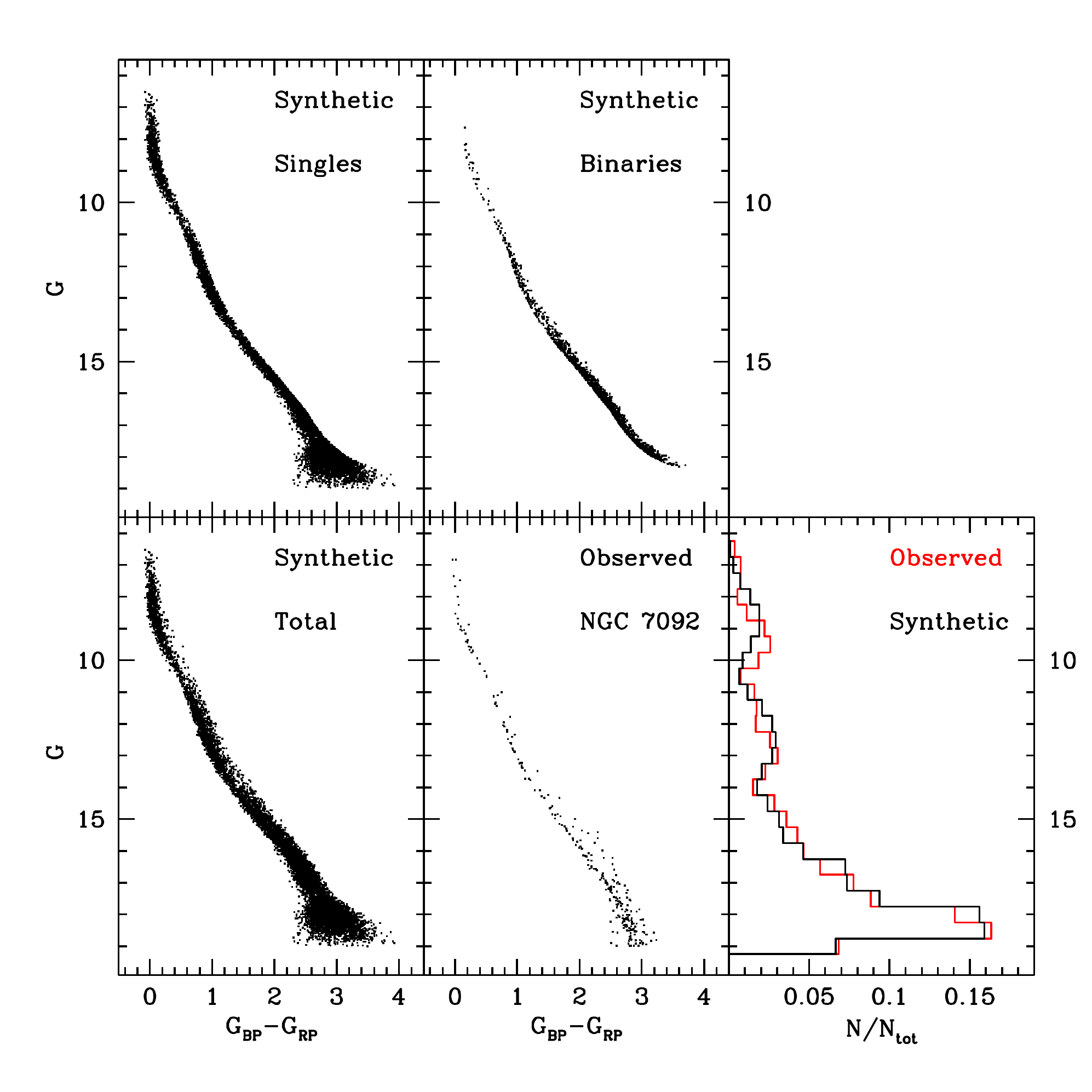}
\caption{Comparison between the best-fit synthetic (bottom-left panel) and observed (bottom-central panel) CMDs for NGC 7092. The CMD of the synthetic single stars (top-left panel) and binaries (top-central panel) are separately shown. The observed (black) and best-fit synthetic (red) $G$-band normalized luminosity functions are compared in the bottom-right panel.} 
\label{syn_obs}
\end{center}
\end{figure*} 

In an iterative algorithm, for a given value of the global binary fraction, all above steps have been repeated and the $k_i$ coefficients have been updated as:
\begin{equation}
k^{\prime}_i=k_i\dfrac{N{^{\text{obs}}_{i}} \sum_{i} N{^{\text{syn}}_{i}}}{N{^{\text{syn}}_{i}} \sum_{i} N{^{\text{obs}}_{i}}},
\end{equation}
where $N{^{\text{obs}}_{i}}$ is the number of single stars (equation \ref{ntot}) and $N{^{\text{syn}}_{i}}$ is the number of the single 
synthetic particles in the corresponding $i-$th bin of the raw MFs. 
In a given iteration, the global binary fraction is also corrected as 
\begin{equation}
f_{b}'=f_{b}\dfrac{N{^{\text{obs}}_\text{bin}} \sum_{i} N{^{\text{syn}}_{i}}}{N{^{\text{syn}}_\text{bin}} \sum_{i} N{^{\text{obs}}_{i}}}
\end{equation}
where $N{^{\text{obs}}_\text{bin}}$ and $N{^{\text{syn}}_\text{bin}}$ are the number of observed (equation \ref{ntot_bin}) and synthetic stars in the binary box, respectively. 
The procedure is continued 
until convergence, providing a best-fit MF and an estimate of the global binary fraction. 

In Fig. \ref{syn_obs}, the comparison between the final synthetic and observed CMDs and $G$-band luminosity functions for NGC 7092 are 
illustrated as an example of the quality of the above procedure.

In the last iteration of the above technique, we evaluate the stellar mass of each OC by normalizing the counted total number of observed single stars ($N{^{\text{sin}}_\text{obs}}$) to those in the best-fitting model ($N{^{\text{sin}}_\text{syn}}$): 
\begin{equation}
M_{\text{lum}}=\frac{N^\text{sin}_\text{obs}}{N^\text{sin}_\text{syn}}\Bigg[\sum_{j_s} m{^{\text{sin}}_{j_s}}+\sum_{j_b}\big(m{^{\text{bin}}_{j_{b1}}}+m{^{\text{bin}}_{j_{b2}}}\big)\Bigg],
\end{equation}
where $m{^{\text{sin}}_{j_s}}$ is the mass of the $j_s-$th single star and $m{^{\text{bin}}_{j_b}}$ is the mass of the $j_b-$th binary. 
In the second term of the right-hand side of this equation, the mass of both components of a given binary has been summed.

To add the contribution to the mass of white dwarfs (WDs), we integrate the mass of the high-mass stars using the corresponding MF adopted in the supersolar regime up to 8 $M_{\odot}$. 
Adopting the initial-final mass relation of \cite{kalirai2009}, $m_{\text{WD}}=0.109 \, m + 0.428$, the total mass fraction ($f_{\text{WD}}$) of WDs has been estimated. 
The total mass of the OC has then been calculated as $M_{\text{tot}}\simeq M_{\text{lum}}(1+f_{\text{WD}})$. 

The half-mass radius ($r_h$) has been derived for each OC by interpolating across its cumulative mass distribution. 
The half-mass relaxation time is then estimated as \citep{spitzer1987}: 
\begin{equation}
t_{rh}=0.138\dfrac{{M_\text{tot}^{1/2}}r_h^{3/2}}{G^{1/2}\bar{m} \ln(\gamma M_\text{tot} / \bar{m})},
\end{equation}
where $\gamma=0.11$ \citep{giersz1996} and $\bar{m}\equiv \sum_i{m_iN_i}/\sum_i{N_i}$ is the mean mass of stars. 
The Jacobi radius of each OC is also estimated as:
\begin{equation}
r_J=R_G^{2/3}\bigg(\frac{GM_\text{tot}}{2V_G^2}\bigg)^{1/3},
\end{equation}
where $R_G$ is the Galactocentric distance of each OC and $V_G=220 \, \text{km/s}$ is the circular velocity of the Galaxy \citep{binney2008}.

\begin{table*}
    \centering
    \caption{Parameters of the best-fitting models: Initial number of the selected stars (2nd column), the total number of member stars after decontamination (3rd column), total mass after correction for fraction of WDs (4th column), half-mass radius (5th column), binary fraction (6th column), half-mass relaxation time (7th column), and Jacobi radius (8th column).}
    \label{table_parameter}
    \begin{tabular}{cccccccccc}  
    \hline     
    Cluster & $N_{\text{ini}}$ & $N_{\text{tot}}$  & $\log(M_{\text{tot}}/{\text{M}}_{\odot})$ & $r_h$ & $f_b$ & $t_{rh}$ & $r_J$ \\
    & & & & (pc) & (Per cent) & (Myr) & (pc)\\
    \hline
    Hyades         & 689  & 582  & 2.812 $\pm$ 0.114 & 3.97 $\pm$ 0.15 & 38 $\pm$ 7 & 151 $\pm$ 35 & 12.51 $\pm$ 1.09 \\
    Coma Berenices & 248  & 174  & 2.167 $\pm$ 0.057 & 4.04 $\pm$ 0.65 & 54 $\pm$ 5 & 117 $\pm$ 31 & 7.61  $\pm$ 0.33 \\
    Pleiades       & 1509 & 1448 & 2.907 $\pm$ 0.026 & 2.96 $\pm$ 0.09 & 27 $\pm$ 3 & 140 $\pm$ 9  & 13.54 $\pm$ 0.27 \\
    IC 2391        & 289  & 255  & 2.207 $\pm$ 0.103 & 1.52 $\pm$ 0.15 & 22 $\pm$ 4 & 29  $\pm$ 7  & 7.83  $\pm$ 0.62 \\
    IC 2602        & 326  & 312  & 2.326 $\pm$ 0.170 & 1.96 $\pm$ 0.17 & 40 $\pm$ 7 & 47  $\pm$ 16 & 8.55  $\pm$ 1.12 \\
    $\alpha$-Per   & 830  & 729  & 2.699 $\pm$ 0.129 & 4.39 $\pm$ 0.13 & 17 $\pm$ 2 & 181 $\pm$ 46 & 11.57 $\pm$ 1.15 \\
    Praesepe       & 1005 & 945  & 2.888 $\pm$ 0.050 & 3.06 $\pm$ 0.12 & 57 $\pm$ 9 & 142 $\pm$ 16 & 13.36 $\pm$ 0.51 \\
    NGC 2451A      & 475  & 401  & 2.418 $\pm$ 0.118 & 3.16 $\pm$ 0.15 & 40 $\pm$ 9 & 110 $\pm$ 26 & 9.26  $\pm$ 0.84 \\
    Blanco 1       & 757  & 657  & 2.594 $\pm$ 0.038 & 3.21 $\pm$ 0.17 & 65 $\pm$ 8 & 162 $\pm$ 17 & 10.51 $\pm$ 0.31 \\
    NGC 6475       & 1706 & 1315 & 3.079 $\pm$ 0.098 & 3.57 $\pm$ 0.13 & 36 $\pm$ 2 & 150 $\pm$ 30 & 14.93 $\pm$ 1.12 \\
    NGC 7092       & 364  & 272  & 2.438 $\pm$ 0.075 & 4.46 $\pm$ 0.24 & 57 $\pm$ 9 & 168 $\pm$ 28 & 9.37  $\pm$ 0.54 \\
    NGC 6774       & 221  & 146  & 2.643 $\pm$ 0.058 & 2.76 $\pm$ 0.35 & 64 $\pm$ 8 & 58  $\pm$ 13 & 10.70 $\pm$ 0.48 \\
    NGC 2232       & 301  & 248  & 2.281 $\pm$ 0.212 & 2.82 $\pm$ 0.24 & 39 $\pm$ 7 & 76  $\pm$ 32 & 8.47  $\pm$ 1.38 \\
    NGC 2547       & 779  & 396  & 2.517 $\pm$ 0.136 & 2.30 $\pm$ 0.18 & 31 $\pm$ 4 & 55  $\pm$ 16 & 9.98  $\pm$ 1.04 \\
    NGC 2516       & 2308 & 1903 & 3.296 $\pm$ 0.099 & 4.62 $\pm$ 0.26 & 47 $\pm$ 8 & 248 $\pm$ 54 & 18.04 $\pm$ 1.37 \\
    \hline  
    \end{tabular} 
\end{table*}
  

\begin{table}
    \centering
    \caption{MF slope of the best-fitting models (third column) in the adopted mass range (second column).}
    \label{table_mf}
    \begin{tabular}{cccccccccc}  
        \hline     
        Cluster & Mass range & $\alpha$  \\
                & ($\text{M}_{\odot}$) &      \\
        \hline
        Hyades         & 0.2 -- 2.5  & -1.62 $\pm$ 0.18 \\
        Coma Berenices & 0.2 -- 2.5  & -1.57 $\pm$ 0.26 \\
        Pleiades       & 0.2 -- 4.9  & -2.61 $\pm$ 0.14 \\
        IC 2391        & 0.2 -- 6.0  & -2.21 $\pm$ 0.13 \\
        IC 2602        & 0.2 -- 6.5  & -2.31 $\pm$ 0.21 \\
        $\alpha$-Per   & 0.2 -- 5.6  & -2.26 $\pm$ 0.15 \\
        Praesepe       & 0.2 -- 2.4  & -2.16 $\pm$ 0.13 \\
        NGC 2451A      & 0.2 -- 6.1  & -2.19 $\pm$ 0.17 \\
        Blanco 1       & 0.2 -- 4.1  & -2.92 $\pm$ 0.22 \\
        NGC 6475       & 0.2 -- 3.1  & -2.26 $\pm$ 0.22 \\
        NGC 7092       & 0.2 -- 3.1  & -2.26 $\pm$ 0.29 \\
        NGC 6774       & 0.3 -- 1.6  & -0.62 $\pm$ 0.23 \\
        NGC 2232       & 0.2 -- 6.5  & -2.72 $\pm$ 0.18 \\
        NGC 2547       & 0.2 -- 5.1  & -2.58 $\pm$ 0.24 \\
        NGC 2516       & 0.3 -- 3.3  & -2.94 $\pm$ 0.21 \\
        \hline  
    \end{tabular} 
\end{table}
  
\section{Results}
\subsection{An overview of OCs' MF and their dynamical parameters}

The estimated dynamical parameters of the 15 OCs of our sample are listed in 
Table \ref{table_parameter}. Our sample 
covers wide ranges in mass ($150\lesssim M/{\text{M}}_{\odot} \lesssim 2000$), 
half-mass radius ($1.5\lesssim r_h/\text{pc} \lesssim 4.7$), half-mass 
relaxation time ($30\lesssim t_{rh}/\text{Myr} \lesssim 250$), and binary 
fraction ($15\lesssim f_b/\text{per cent} \lesssim 65$).

\begin{figure}
\begin{center}
\includegraphics[height=18cm, width=85mm]{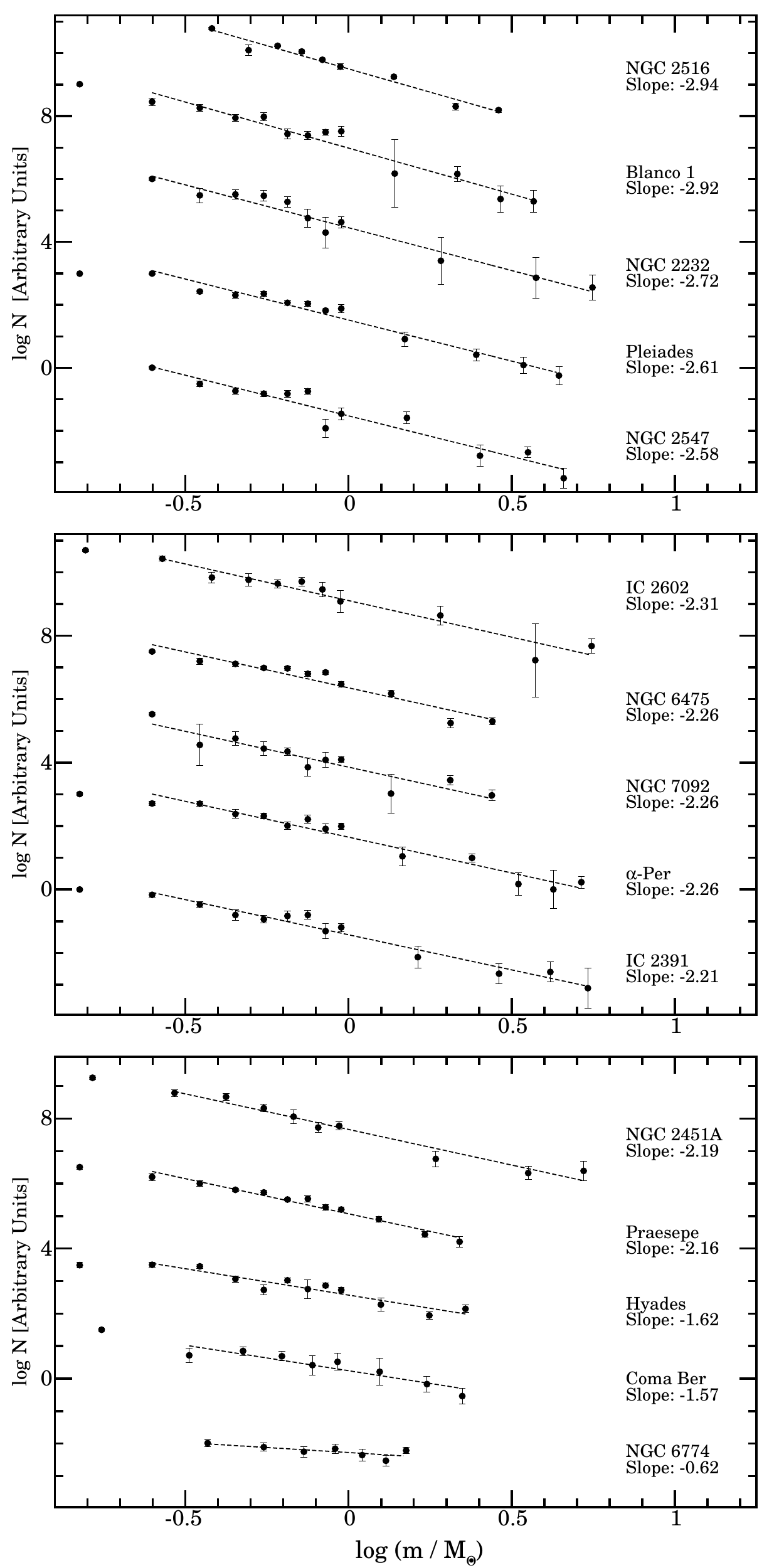}
\caption{PDMF of 15 OCs in our sample. A vertical shift has been added to each cluster in each panel for clarity. From top to bottom panels, PDMFs of OCs have been sorted from the steepest MF to the flattest one. The dashed lines represnt the corresponding linear best-fit to the MF of the OCs.}
\label{mf}
\end{center}
\end{figure}

\begin{figure}
\begin{center}
\includegraphics[height=80mm, width=85mm]{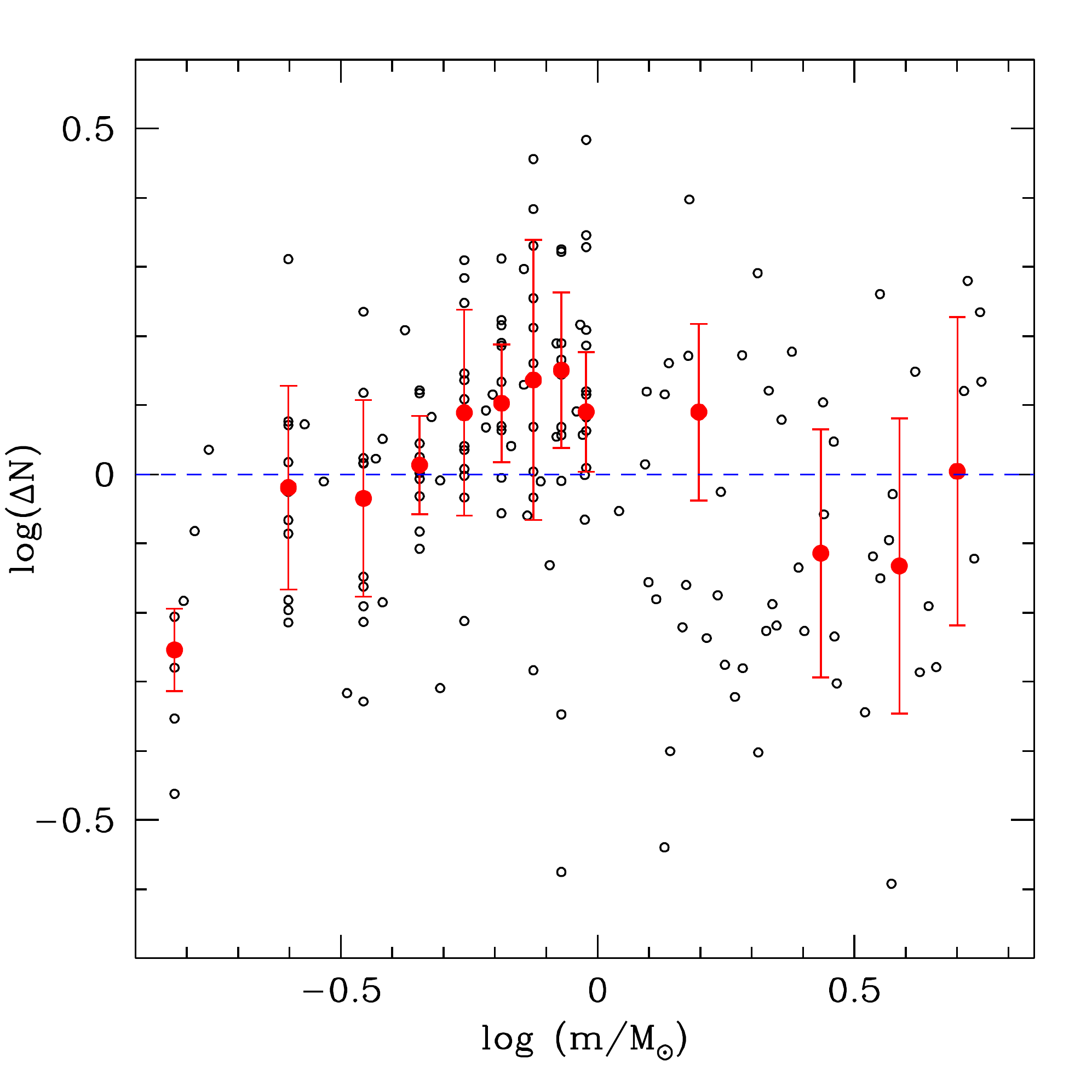}
\caption{Residuals of the power-law fit for the whole OCs sample (black circles). The average residuals and their standard deviations for all mass bins are marked with red circles and error bars.} 
\label{residual}
\end{center}
\end{figure} 

The derived PDMF of the 15 OCs are shown in Fig. \ref{mf} and 
their global MF slopes are listed in Table \ref{table_mf}. 
The PDMF of our OCs sample widely varies from 
the steepest MF with $\alpha=-2.94$ for NGC 2516 to the flattest one with 
$\alpha=-0.62$ for NGC 6774. 

In Fig. \ref{residual}, the residuals of the 
power-law fit for all 15 OCs are shown. It is clear that the residuals have a 
convex shape with a systematic deviation at $m<0.2 \, {\text{M}}_{\odot}$ 
(residual > 0.2) and a bend in the subsolar regime. 
This evidence in young poorly evolved clusters could indicate a deviation 
of the IMF from a simple power-law function \citep{baumgardt2017}, in agreement with 
the prediction of star formation theories \citep{adams1996,hennebelle2012}.
Similar behaviour has been also shown by 
\cite{sollima2017} for GCs.  

\subsection{Comparison with Other Works}

In Table \ref{table_comparison}, we list the derived MF slopes of various 
studies which analysed samples of OCs with an overlap with our analysis. 
We mentioned only the 
studies which reported the evaluated power-law MF giving the exact values of 
slope and covering partially or completely our adopted mass ranges. 
Notice that most studies have divided the MFs into two or more segments while 
our derived single-slope MFs cover the wide mass ranges.

Our derived MF slopes are reasonably in good agreement with those reported 
in the following studies within $\Delta \alpha\simeq\pm 0.2$: MF of NGC 2516 corrected for binaries by 
\cite{sung2002} in the $0.8<m/\text{M}_\odot<5$ mass range, MF of NGC 2547 \citep[using models of ][]{d'antona1997} reported by \cite{naylor2002} in the $1<m/\text{M}_\odot<6$ mass range, MFs of $\alpha$-Per derived by 
\cite{sheikhi2016} within $0.62<m/\text{M}_\odot<4.68$, MF of IC 2391 evaluated by 
\cite{sanner2001} within $1.1<m/\text{M}_\odot<8.1$, MFs of Pleiades 
reported by \cite{moraux2003} and \cite{roser2020} in the $1.5<m/\text{M}_\odot<10$ 
and $1<m/\text{M}_\odot<2.5$ mass ranges, respectively.

Our derived MF slopes are flatter than those predicted by the 
following papers in the mass range covering partially or completely the 
supersolar mass regime: \cite{roser2011} and \cite{goldman2013} for Hyades, 
\cite{bouy2015}, \cite{sollima2019}, and \cite{niu2020} for Pleiades, 
\cite{sanner2001} for Pleiades, $\alpha$-Per, Blanco 1, and NGC 7092, 
\cite{khalaj2013} for Praesepe, \cite{naylor2002} for NGC 2547 \citep[using models of][]{siess2000}. 
Our derived MFs are instead steeper than the literature estimates for the rest of the cases listed in Table \ref{table_comparison}. 
A possible reason for this discrepancy can be attributed to the 
different mass ranges covered by the various studies: while our adopted mass 
range is relatively wide, the flattest MFs reported in the studies listed in Table 
\ref{table_comparison} cover only the subsolar regime, including even 
substellar sources down to brown dwarfs, e.g. \cite{barrado2004} for IC 2391 and 
\cite{moraux2007} for Blanco 1. Besides this effect, the differences in the predicted MFs can be 
the results of adopting different datasets, isochrones, ages and metallicities 
for each OC \citep{ebrahimi2020}.                    

The global binary fractions in our OCs sample widely vary between 17 and 65 
per cent and are compatible with those derived by \cite{bica2005} for 9 OCs 
($15<f_b/\text{per cent}<54$), \cite{sharma2008} ($30<f_b/\text{per cent}<75$) 
for 9 OCs, \cite{sollima2010} in the core of 5 OCs ($35<f_b/\text{per cent}<70$), 
\cite{cordoni2018} for 12 OCs ($11<f_b/\text{per cent}<51$), \cite{niu2020} for 
12 OCs ($13<f_b/\text{per cent}<47$), and \cite{jadhav2021} for 23 OCs 
($30<f_{b,\text{mean}}/\text{per cent}<53$).

The Roche-lobe filling factors in our OCs sample cover the range
$0.15\lesssim r_h/r_J \lesssim 0.55$ in agreement with those estimated by 
\cite{angelo2019} for 12 OCs in the Sagittarius spiral arm 
($0.15\lesssim r_h/r_J \lesssim 0.45$) and by \cite{angelo2020} for 27 Galactic 
OCs ($0.2\lesssim r_h/r_J \lesssim 0.7$). If we assume $r_h/r_J \simeq 0.07$ as 
a boundary for distinguishing underfilling clusters from filling ones 
\citep{baumgardt2010}, all OCs in our sample are strongly tidally filling.

\begin{figure*}
\begin{center}
\includegraphics[width=150mm]{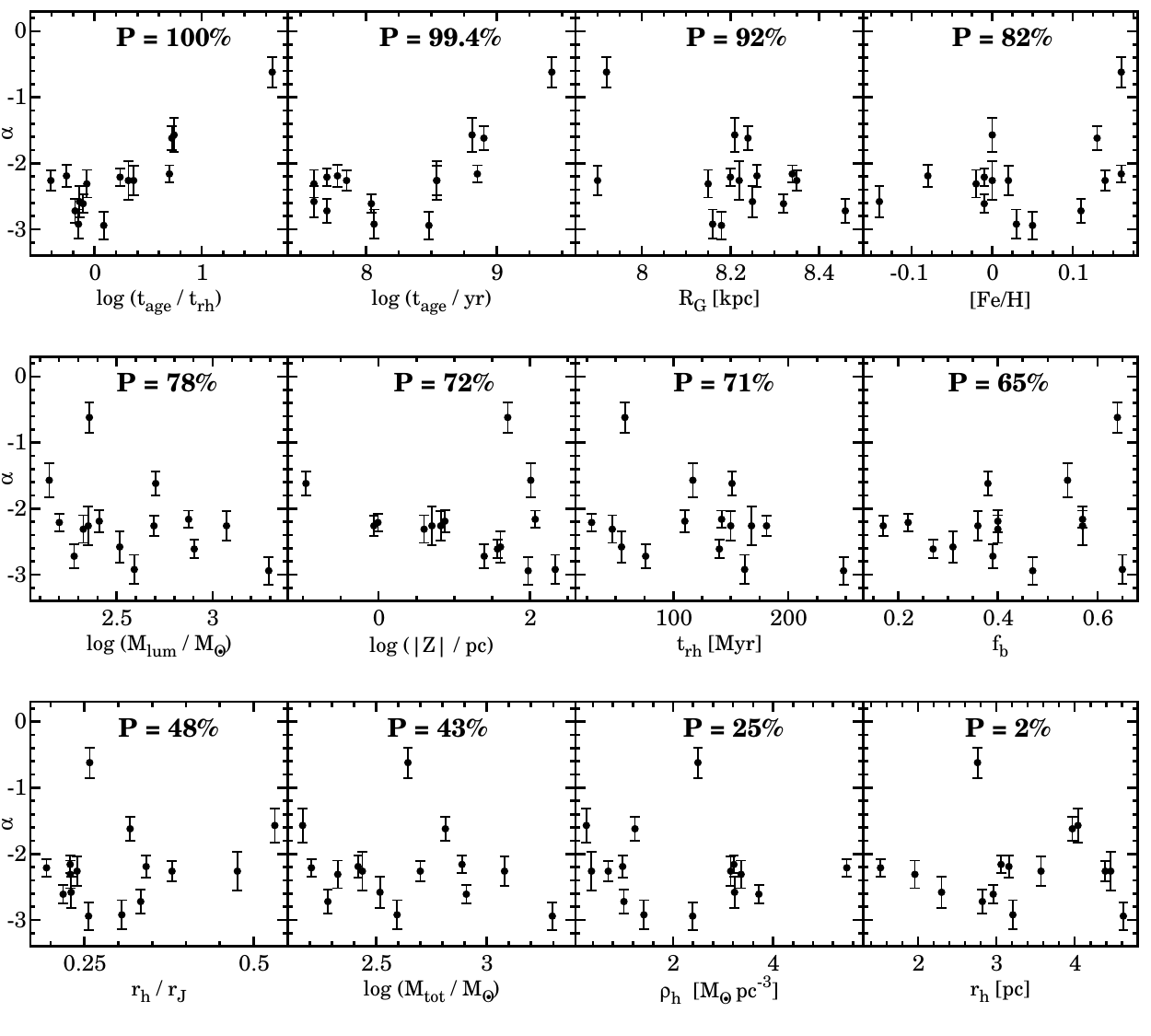}
\caption{Univariate correlations between the global PDMF slope ($\alpha$) and various cluster parameters. The statistical significance ($P$) of each correlation is shown.}
\label{correlation}
\end{center}
\end{figure*}

\subsection{Analysis of correlations} \label{Sec:4.2}

The possible correlations between PDMF slope and the structural and dynamical 
parameters of our OCs sample is investigated in this section. 
The following parameters have been considered: the stellar mass 
($M_{\text{lum}}$), the total mass ($M_{\text{tot}}$), 
the half-mass radius ($r_h$), the half-mass relaxation time ($t_{rh}$), 
the binary fraction ($f_b$), the half-mass density 
($\rho_h\equiv 3M_{\text{tot}}/[8\pi r_h^3]$), and the Roche-lobe filling factor 
($r_h/r_J$) (all estimated in the present work), the age ($t_{\text{age}}$) and 
the metallicity [Fe/H], derived by \cite{gaia2018} and \cite{pang2021},
the Galactocentric distance ($R_G$) and the distance from Galactic plane $Z$ (using 
the OCs positions in Galactic coordinates derived from 
WEBDA\footnote{https://webda.physics.muni.cz/}) and the ratio of age to 
half-mass relaxation time ($t_{\text{age}}/t_{rh}$).   

We adopt a permutation test to investigate the univariate correlations 
between PDMF slope and the other parameters and calculate the statistical 
significance probability ($P$) for each correlation \citep{ebrahimi2020}. 
In Fig. \ref{correlation}, the entire set of correlations and their 
corresponding probabilities are shown. A tight significant correlation 
($P>99.7 \, \text{per cent}$) between $\alpha$ and $\log(t_{\text{age}}/t_{rh})$ 
and a marginal correlation ($95<P/ \text{per cent}<99.7$) with $t_{\text{age}}$ 
are found. The other parameters show no significant correlation with the PDMF 
slope ($P< 95 \, \text{per cent}$).

According to the distribution of OCs in $\log(t_{\text{age}}/t_{rh})-\alpha$ 
plane, we can classify the OCs into three groups:
\begin{enumerate}
\item{Less evolved OCs, younger than their relaxation time, with a PDMF 
with $-3 \lesssim \alpha \lesssim -2.2$ which is compatible or even 
steeper than the \cite{salpeter1955} IMF ($\alpha_{\text{Sal}}=-2.35$). 
The OCs in this group occupy the largest portion in 
$\log(t_{\text{age}}/t_{rh})-\alpha$ plane.}
\item{Moderately evolved OCs, with an age of the same order of their $t_{rh}$, with a PDMF ($-2.2 \lesssim \alpha \lesssim -1$) flatter than Salpeter IMF.}
\item{An extremely evolved OC (NGC 6774), with $t_{age}> 10\,t_{rh}$ with the 
flattest PDMF ($\alpha>-1$), enhancing the correlation between PDMF slope and 
$t_{\text{age}}/t_{rh}$.} 
\end{enumerate} 

Various works tried to investigate the correlation between PDMF slope and 
$t_{\text{age}}/t_{rh}$. \cite{bonatto2005} estimated the global PDMF of 11 
nearby open clusters and showed that the PDMF of low-mass and high-mass OCs 
noticeably flatten when their ages become older than $\simeq 7~t_{rh}$ and 
$\simeq 11~t_{rh}$, respectively. \cite{maciejewski2007} performed similar 
work for an extended sample of 42 OCs and found that this flattening occurs for 
OCs with $t_{\text{age}}\gtrsim 100\,t_{rh}$ and the PDMF of the other OCs with
 $t_{\text{age}}\lesssim 100\,t_{rh}$ are generally close to the \cite{salpeter1955} 
 IMF. \cite{sharma2008} have also estimated the overall PDMF of 9 OCs and 
 indicated that the most evolved OCs have the flattest PDMF. 

\cite{piatti2019} reported an anticorrelation between the Roche-lobe 
filling factor ($r_h/r_J$) and $t_{\text{age}}/t_{rh}$ for a sample of 12 OCs 
while we did not find any significant correlation between these two quantities.  

To test the most significant correlations between parameters, we expand our 
analysis to bivariate correlation using the Monte Carlo method proposed 
by \cite{sollima2017}. We found that if we assume $t_{\text{age}}/t_{rh}$ as 
the first independent variable, it does not exist any other parameter providing a 
significant bivariate correlation. Instead, If we adopt any other parameter as 
the first independent variable, the only significant bivariate correlation is 
with $t_{\text{age}}/t_{rh}$ as the second independent variable. This fact implies 
that two-body relaxation is the dynamical process which governs the shape of the PDMF.

\subsection{Comparison Between the Evolution of OC and GC MFs} \label{Sec:4.3}

Recently, we have analysed the deep photometric data of 31 Galactic GCs 
observed with the Wide Field Channel of the Advanced Camera for Surveys of the 
\emph{Hubble Space Telescope} \citep{ebrahimi2020} and confirmed that the GCs' 
PDMF slope in $0.2<m/\text{M}_{\odot}<0.8$ range correlates with 
$t_{\text{age}}/t_{rh}$. As reported in Section \ref{Sec:4.2}, a similar result 
has been obtained for our sample of 15 Galactic OCs. 
Therefore, $t_{\text{age}}/t_{rh}$ is the most significant dynamical driver of 
the MF slope evolution for 
both the old GCs ($t_{\text{age}}>11 \, \text{Gyr}$) and young OCs 
($t_{\text{age}}<3 \, \text{Gyr}$).

\begin{figure}
\begin{center}
\includegraphics[width=85mm]{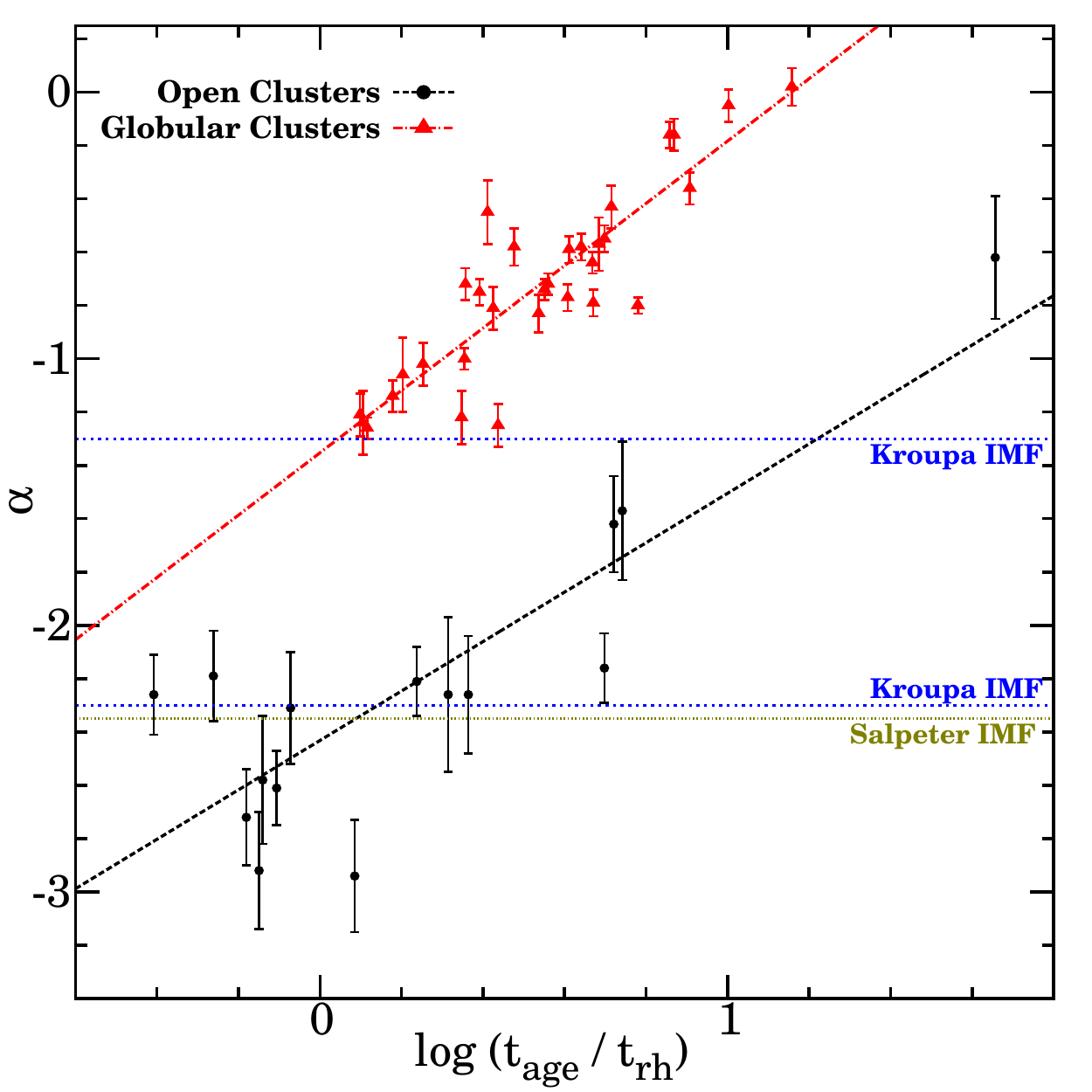}
\caption{Slope of GC (red dots) and OC (black dots) PDMFs versus the ratio of age to present-day half-mass relaxation time. 
The least-square fits on OCs and GCs samples are marked by the dashed black and dotted-dashed red lines, respectively. 
The IMF slopes of \protect\cite{salpeter1955} and \protect\cite{kroupa2001} are represented by horizontal green and blue dotted lines.}
\label{oc_gc}
\end{center}
\end{figure}

In Fig. \ref{oc_gc}, the evolution of OCs and GCs in the 
$t_{\text{age}}/t_{rh}-\alpha$ plane is shown. It is clear that the OCs and GCs 
are evolving along different, almost parallel, paths.    
In particular, at a given number of elapsed half-mass relaxation times, OCs have MF slopes steeper by $\Delta \alpha\simeq 1$ than GCs.
A linear least-square fit of the two groups gives the following relations.

\begin{equation}
\alpha_{\text{GC}}=(1.17\pm 0.12)\log (t_{\text{age}}/t_{rh})_{\text{GC}}-(1.35\pm 0.07),
\end{equation}
while the similar fit for the OCs sample in the present work is given by 
\begin{equation}
\alpha_{\text{OC}}=(0.93\pm 0.16)\log (t_{\text{age}}/t_{rh})_{\text{OC}}-(2.43\pm 0.09).
\end{equation}

Such a difference is significant ($\Delta \alpha=1.08\pm 0.11$ at $t=t_{rh}$) and can be hardly explained in terms of dynamical evolution.
It is worth noting that the mass range where the GCs MF is measured is generally limited to low-mass ($m<0.8 \, {\text{M}}_{\odot}$) stars, still unevolved in these old stellar systems. Instead, in the present analysis, we adopted a large upper limit. However, the same evidence is apparent even measuring the OCs MF slopes in a reduced mass range $0.2<m/{\text{M}}_{\odot}<0.8$.

\subsection{Mass Segregation} \label{Sec:4.4}

To estimate the degree of mass segregation for each OC in our sample, we 
defined five radial bins containing an equal number of stars. 
The number of stars in each radial bin has been corrected for field contamination using
equation \ref{ntot}, where in this case $A_{\text{in}}$ is the 
corresponding annular area confining each radial interval. The same iterative 
method which is explained in detail in Section \ref{Sec:3} has been employed 
to compute the PDMF slope in all radial bins for each OCs. Since 
observational uncertainties lead to an unrealistic prediction of the radial 
behaviour of the binary fraction in each individual bin, particularly in less populated OCs, 
we assumed that the binary fraction is equal to the global one in all radial 
bins. A least-squares fit in $\log(r)-\alpha$ plane has then been 
performed and the slope of each fit, $\alpha_{R}=\text{d}\alpha/\text{d}[\log(r)]$, 
has been evaluated as an indicator of the degree of mass segregation \citep[see also][]{webb2016}. 
In other words, an OC with a large value of $\alpha_{R}$ appears to be highly mass 
segregated.

\begin{figure}
\begin{center}
\includegraphics[width=82mm]{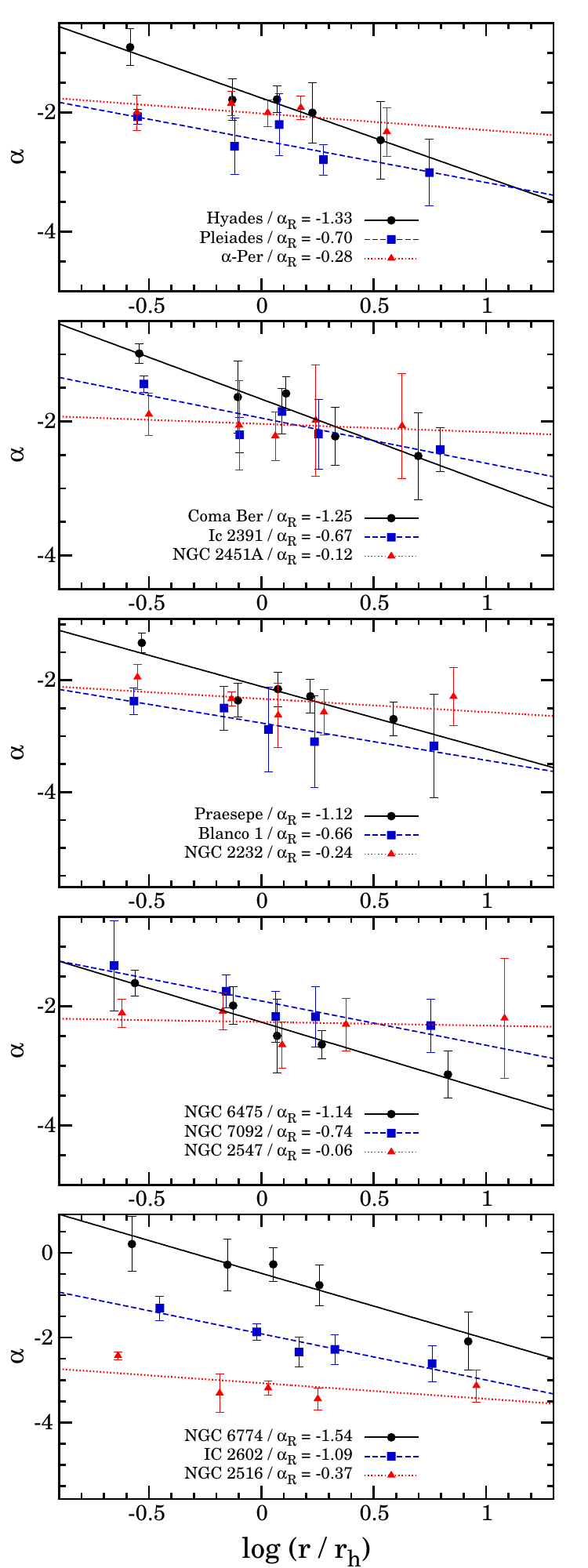}
\caption{PDMF slopes of all 15 OCs measured in 5 different radial bins. The local PDMF slopes of all 15 OCs are marked by dots with different colours and symbols. 
The corresponding linear least-square fits are marked by solid, dotted and dashed lines. The slope of each linear fit ($\alpha_R$) is shown for each OC in each panel.}
\label{mass_seg_slope}
\end{center}
\end{figure}

In Fig. \ref{mass_seg_slope}, the radial variation of the PDMF slopes of the 
entire OCs sample is shown. We can classify 
OCs on the basis of their degree of mass segregation in the following three 
groups:
\begin{enumerate}
\item{OCs with strong mass segregation with $\alpha_R<-1$ (marked with black dots and solid lines in Fig. \ref{mass_seg_slope});}
\item{OCs with intermediate mass segregation with $-1<\alpha_R<-0.5$ (marked with blue dots and dashed lines in Fig. \ref{mass_seg_slope});}
\item{OCs with weak or no evidence of mass segregation with $-0.5<\alpha_R<0$ (marked with red dots and dotted lines in Fig. \ref{mass_seg_slope}).}
\end{enumerate}

\begin{figure}
\begin{center}
\includegraphics[width=85mm]{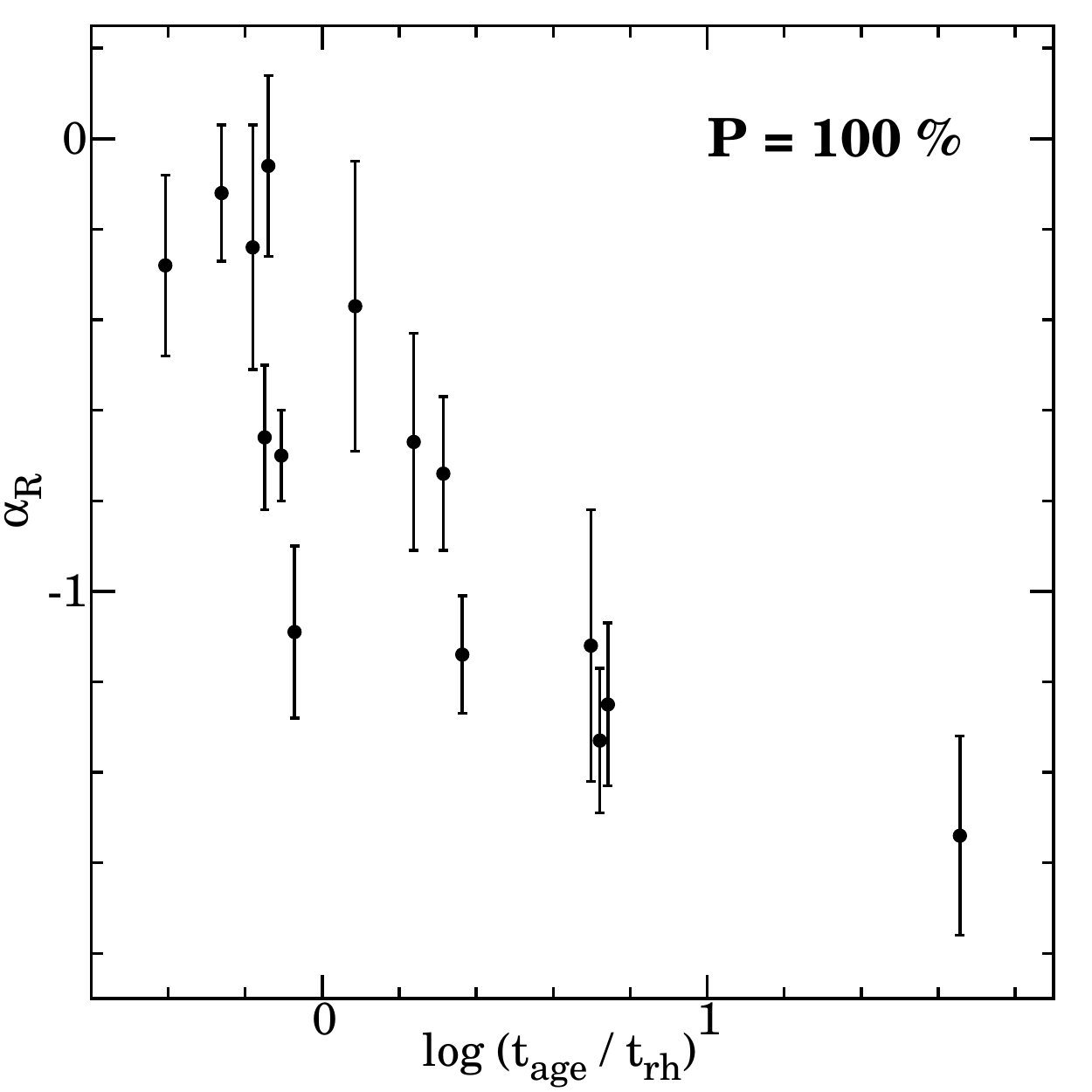}
\caption{Mass segregation parameter ($\alpha_R$) with respect to the ratio of age to present-day half-mass relaxation time. The statistical significance probability ($P$) of the correlation is also shown.}
\label{mass_seg_corr}
\end{center}
\end{figure}

In Fig. \ref{mass_seg_corr}, the relation between the degree of mass segregation 
parameter ($\alpha_R$) and $t_{\text{age}}/t_{rh}$ is illustrated, showing a 
significant correlation between them. This fact implies that the most (less) 
evolved OCs are the most (less) mass segregated. 
This result is in agreement with theoretical expectations \citep{webb2016}.

\cite{sharma2008} estimated 
the PDMF slope of nine OCs separately in the inner and outer parts of them. 
They showed that the PDMF slope differences ($\alpha_{\text{in}}-\alpha_{\text{out}}$; 
adopted as an indicator of the degree of mass segregation) decrease with the 
ratio of age to the relaxation time. \cite{maciejewski2007} estimated the same 
indicator (the difference between the PDMF slope of the core and halo) for 42 
OCs and found no relation between it and $t_{\text{age}}/t_{rh}$. 
However, they noticed that the stronger mass segregation occurs in OCs older 
than their relaxation times.  

\subsection{Comparison with Monte Carlo Simulations} \label{Sec:4.5}

To investigate the possible IMF and its subsequent evolution in OCs, 
we compare our observational PDMF to those of a set of 
Monte Carlo simulations. We used the Monte Carlo code described in \citet{sollima2014} to simulate two models with different sets of 
initial conditions which are summarized in Table \ref{table_mc}. 
The IMF of the first model starts with the \cite{salpeter1955} IMF 
($\alpha_\text{IMF}=-2.3$) and the second one has a steeper IMF 
($\alpha_\text{IMF}=-3.0$). Both simulated clusters are assumed on circular 
orbits in an isothermal halo with circular velocity $V_{G}=220 \, \text{km/s}$ at a Galactocentric distance 
$R_G=8.1 \, \text{kpc}$ to mimic a typical OC in the solar neighbourhood. 
Particles were distributed following a \citet{king1966} profile with central adimensional potential $W_{0}=5$ and core radius $r_{c}=1\,\text{pc}$.
All simulated clusters have initial binary fractions of $f_{b}=50\%$.
They are allowed to evolve until their dissolution time at 
$t_{\text{diss}}\simeq2 \, \text{Gyr}$, matching the maximum lifetime of OCs in 
our sample. The MF at each time step is calculated in the 
$0.2<m/\text{M}_{\odot}<2.0$ mass range adopting 9 and 4 mass bins in subsolar 
and supersolar regimes, respectively. The degree of mass segregation slopes 
($\alpha_R$) is also evaluated by dividing the sample radially into three annuli 
containing the same number of stars. 

\begin{figure}
\begin{center}
\includegraphics[width=85mm]{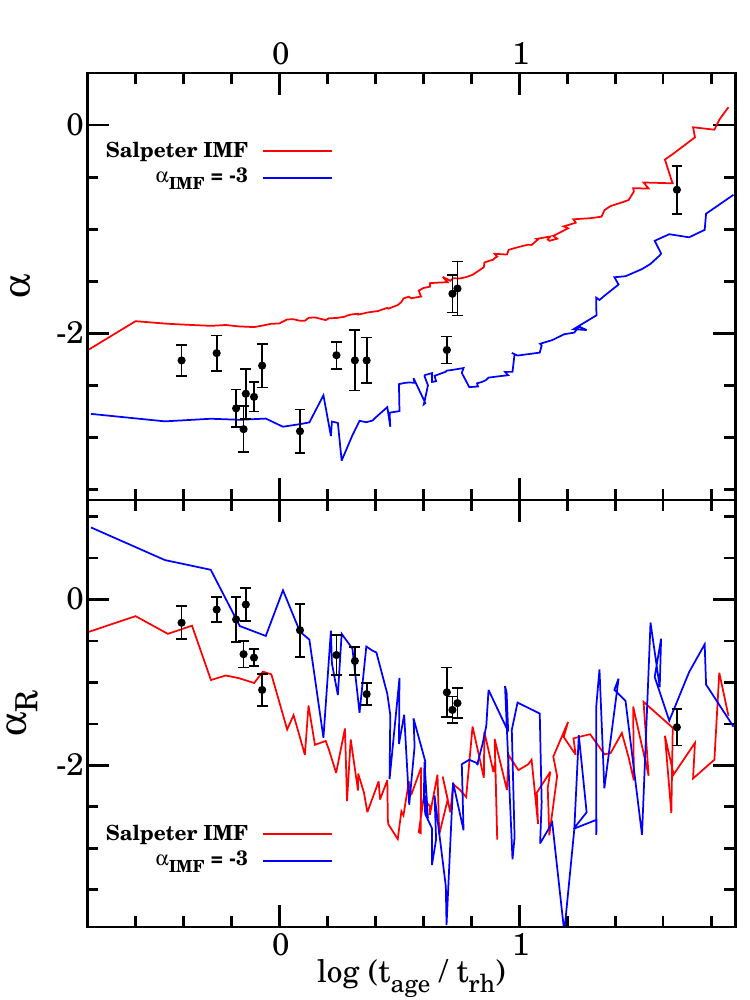}
\caption{Comparison of the PDMF slopes (top panel) and degree of mass segregation slopes ($\alpha_{R}$; bottom panel) as a function of the ratio of age to present-day half-mass relaxation time in our OCs sample (filled points)
and in the Monte Carlo simulations with $\alpha_{\text{IMF}}=-2.3$ (red lines) and $\alpha_{\text{IMF}}=-3$ (blue lines).}
\label{oc_mc}
\end{center}
\end{figure}

In Fig. \ref{oc_mc}, the evolution of the PDMF and $\alpha_R$ of these two 
simulations are presented along with observational data. 
Qualitatively, the trend of the evolutionary tracks is in good agreement with 
observational data. In particular, the MF remains constant at $\log(t/t_{rh})<0.3$ and tends to flatten at later epochs as a result of two-body relaxation.
The tracks of simulations with different IMF maintain an almost constant MF slope difference along the entire evolution.
Observed OCs lie in the region comprised between these two tracks, suggesting that their IMFs were all in the range $-3<\alpha<-2.3$. 
According to the bottom panel of Fig. \ref{oc_mc}, in spite of the large noise due to number counts fluctuations, OCs evolve
toward an increasing degree of mass segregation, in agreement with the N-body simulations of \citet{webb2016}.       

\begin{table}
    \centering
    \caption{Initial dynamical parameters of the Monte Carlo simulations.}
    \label{table_mc}
    \begin{tabular}{cccccccccc}  
        \hline     
        Model & $N_0$ & $\alpha_{\text{IMF}}$ & $M_0$                & $r_{h0}$ & $f_{b0}$   \\
              &       &                       & ($\text{M}_{\odot}$) & (pc)     & (per cent) \\
        \hline
        1 & 5523 & -2.3  & 3037 & 2.00 & 50  \\
        2 & 3336 & -3.0  & 1011 & 2.00 & 50  \\
        \hline  
    \end{tabular} 
\end{table}  

\subsection{Comparison Between the PDMF of OCs and the Solar Neighbourhood} \label{Sec:4.6}

Assuming that the embedded star clusters are the building blocks of galaxies, most of the stars in the solar neighbourhood as well as the Galactic disc, forms from the dissolution of young associations and clusters through the process of rapid gas expulsion and subsequent unbinding and expansion of the embedded cluster \citep{KroupaAH2001, kruijssen2012, Brinkmann2017}. 
Therefore, it is interesting to compare the MF of our OCs with that of the solar neighbourhood.

\cite{sollima2019} has computed 
the solar neighbourhood IMF sampling more than 120000 field stars in the 
Galactic disc, including the effects of unresolved binary fraction, 
distribution of the metallicity, and star formation history. He found the 
related IMF as a two-segmented power-law function with slopes $\alpha_1=-1.34$ 
in the subsolar and $\alpha_2=-2.41 \, \text{to} -2.68$ in the supersolar regime, according to the adopted set of isochrones. 
Similar results have been obtained by \citet{hallakoun2021} from the analysis of the same dataset.

We interpolated the MF of the solar neighbourhood by \citet{sollima2019} at the mass 
bins ($m_i$) of a given OC and calculated the difference between the two MFs ($\Delta N_i$) at each i-th mass bin. 
The slope ($\alpha_{\text{res}}$) of a least-squares fit in the $\log(m_i)-\Delta N_i$ 
plane is then derived for each OC.

\begin{figure}
\begin{center}
\includegraphics[width=85mm]{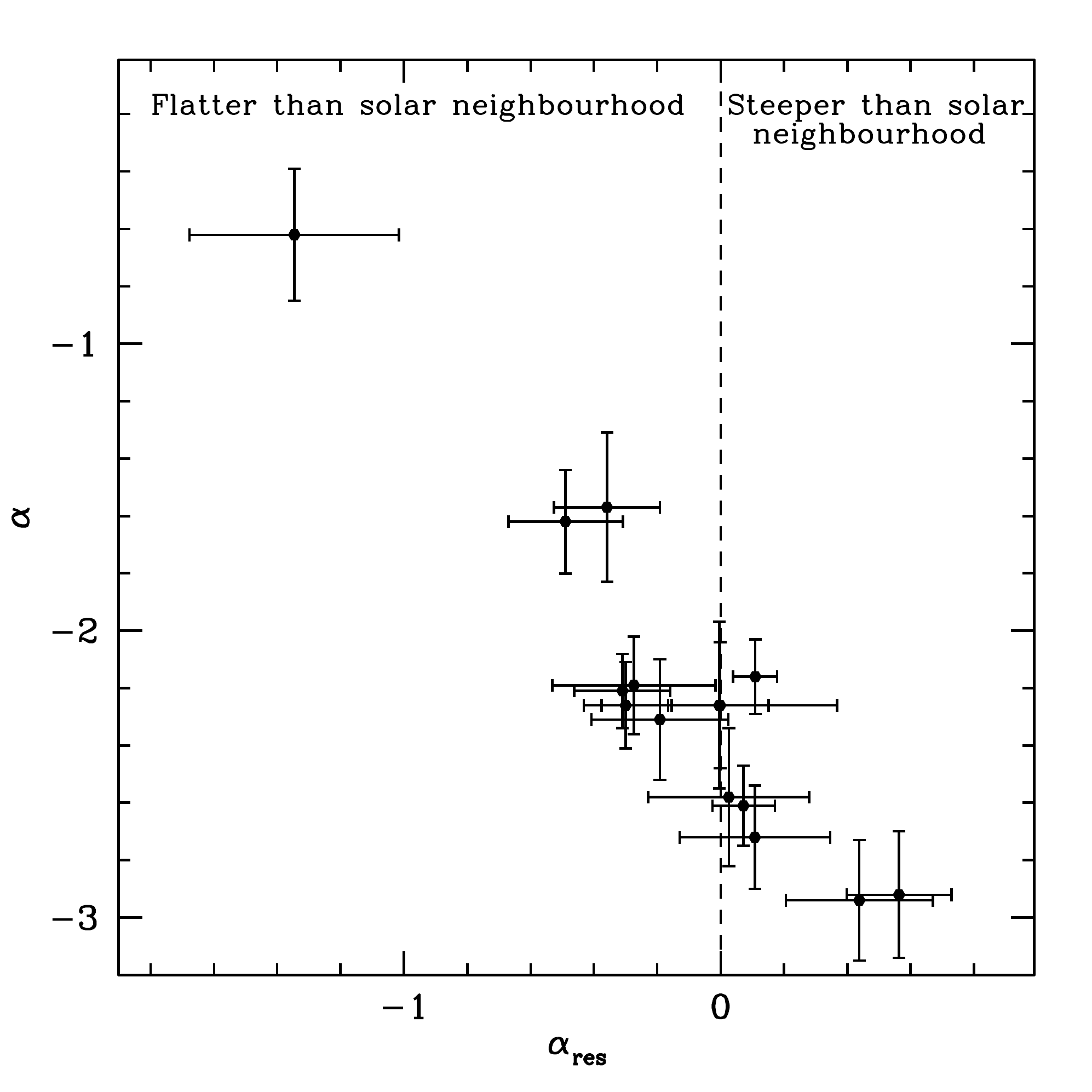}
\caption{Slopes of MF residuals with respect to the solar neighbourhood IMF as a function of the PDMFs slope. The dashed line marks the zero difference locus.}
\label{res_sn}
\end{center}
\end{figure}

The residual slope ($\alpha_{\text{res}}$) and its associated uncertainty 
indicate the similarity between these two MF slopes. Thus, for an OC with a MF 
steeper than the solar neighbourhood, the residual slope is $\alpha_{\text{res}}>0$ 
and it is $\alpha_{\text{res}}<0$ when it has a flatter MF. 

In Fig. \ref{res_sn}, the values of $\alpha_{\text{res}}$ for all 15 OCs are 
shown against their PDMF slopes. It can be seen that, while 
the evolved OCs with $\alpha>-1.7$ (i.e. NGC 6774, Hyades, and Coma 
Berenices) have $\alpha_{\text{res}}\ll 0$, all the other clusters have residual slopes close to zero. 
A $\chi^2$ test 
indicates a probability of $P_{\chi^2}=99.7$ per cent that all the 12 less evolved OCs have the 
same MF slope of the solar neighbourhood. This finding is consistent with the 
scenario linking the OCs dissolution with the origin of the Galactic disc 
population.    

\section{Conclusion}

In this paper, we used the high quality \emph{Gaia} EDR3 database to select 
the member stars of 15 Galactic OCs located in the solar neighbourhood, 
thanks to their distribution in a 5D parameter space ($G$, $G_{BP}-G_{RP}$, $\mu^*_{RA}$, 
$\mu_{Dec}$, $p$). 
We performed an iterative technique to find a best-fit synthetic stellar population reproducing the 
observational CMD for each OC. 
The global PDMF, binary fraction and dynamical parameters have then been 
determined for all the OCs of the sample. 

The derived PDMFs of the 15 OCs vary in the range 
($-3\lesssim \alpha \lesssim -0.6$) and cover less to most evolved OCs with 
$0.4\lesssim t_{\text{age}}/t_{rh}\lesssim 40$. We estimated the binary fraction 
in our OCs sample, ranging between 15 to 65 
per cent, which is generally comparable with those estimated in previous studies 
\citep{sollima2010, cordoni2018, niu2020}. 

By analysing the univariate correlations between the structural and dynamical 
parameters, we found a significant correlation between the PDMF slope and the ratio 
of age to present-day half-mass relaxation time.  
A marginal correlation with the age of OCs has been also confirmed, while no 
significant correlation with any other parameter was found. 
The investigation of the bivariate correlations implies that the 
$t_{\text{age}}/t_{rh}$ is the only significant parameter which regulates the 
shape of MF in OCs. Moreover, we showed that all OCs in our sample are tidally 
Roche-lobe filling clusters with $r_h/r_J>0.15$. These observational facts are 
consistent with the theoretical paradigm predicting that the 
complex combination of the internal processes like two-body relaxation and 
external mechanisms, e.g. mass loss due to Galactic tidal field, are 
responsible for the evolution of the OCs structure and MF. 

The same conclusion has been found for 
a large sample of Galactic GCs by \cite{sollima2017}, \cite{baumgardt2018}, 
and \cite{ebrahimi2020}, indicating that the same mechanisms drive the 
evolution of both young and old star clusters. However, we found that the 
evolutionary tracks of OCs and GCs are different. Considering that the main physical 
process determining the preferential evaporation of low-mass stars is two-body relaxation 
\citep{spitzer1969}, one would expect that stellar systems born with the same IMF 
should evolve in similar ways after the same number of elapsed half-mass relaxation times \citep[see e.g.][]{baumgardt2003}.  
So, the different paths followed by OCs and GCs in the $t/t_{rh}-\alpha$ plane suggest different IMFs for these stellar systems.
In this regard, OCs and GCs are part of different Galactic components: while the 
formers were born in the metal-rich Galactic disc during its dissipative collapse, the latter objects are part of the metal-poor halo and can possibly form in external satellites \citep{massari2019}.
Consider that a dependency of the MF slope on the metallicity has been proposed by many authors 
\citep{portinari2004,ballero2006,hoversten2008,vandokkum2010,geha2013,dib2014,ebrahimi2020} and predicted theoretically by \cite{silk1977}, \cite{adams1996}, \cite{hennebelle2012}, and \cite{marks2012}. 

By estimating the degree of radial variation of the MF slope
($\alpha_R$), we investigated the degree of mass segregation for each OC. 
We found a group of OCs with strong mass segregation ($\alpha_R<-1$), e.g. 
NGC 6774, Coma Berenices, and Hyades, while others appear not to be mass 
segregated ($\alpha_R\simeq 0$), e.g. NGC 2547 and NGC 2451A. 
Moreover, we found a significant correlation between $\alpha_R$ and 
$t_{\text{age}}/t_{rh}$, suggesting that the most evolved OCs are more mass 
segregated. This is in agreement with the predictions of the N-body simulations of \citet{webb2016}
who also found that mass segregation increases as two-body relaxation becomes more effective.

The evolution of the PDMF and its radial variation have been simulated by two Monte 
Carlo models and compared with those estimated for all the 15 OCs in this work. 
The evolution of the two simulations with different IMF slopes 
$\alpha_{\text{IMF}}=-2.3$ (Salpeter IMF) and  $\alpha_{\text{IMF}}=-3$ encompass
the locus of the observational OC PDMFs in both the $t/t_{rh}-\alpha$ and $t/t_{rh}-\alpha_{R}$ 
planes and suggest that all OCs of our sample could have been born with IMF slopes $-3<\alpha<-2.3$.

The PDMF of all OCs in our sample was compared with that of the solar 
neighbourhood estimated by \cite{sollima2019} using the \emph{Gaia} database. 
Besides a group of 3 evolved OCs, the slopes of MF residual are 
around zero, implying that the PDMFs of OCs are similar to that of the solar 
neighbourhood. This fact supports the
hypothesis postulating that the dissolution of the OCs is at the origin of the Galactic 
disc population \citep{kruijssen2012}.

The present study is an attempt to investigate the present-day dynamical 
properties of Galactic OCs using the newest data released by the \emph{Gaia} 
mission. The unprecedented quality of the astrometric solutions of 
\emph{Gaia} EDR3, allowed us to distinguish the members of OCs from the 
background/foreground contaminations better than in previous studies. 
However, the completeness limits in \emph{Gaia} photometric data for 
faint/low-mass stars restricted us to analyze only OCs with distances less than 
420 pc from the Sun. The unclear identification of very young gravitationally bound
systems from the embedded cluster in OB associations, the large biases in 
determining the dynamical properties of less-populated OCs, and the effect of 
interstellar reddening variations on the CMD of the OCs are the other limiting 
factors in selecting the target OCs \citep{gaia2018}. The investigation of an 
extended sample of the Galactic OCs using the updated datasets and developed 
techniques will be the next steps in our knowledge about dynamical properties 
of the OC in the future. 

\section*{Acknowledgments}

HE acknowledges the support under a postdoctoral research agreement between Iran National Science Foundation (INSF) and Institute for Advanced Studies in Basic Sciences (IASBS) through the grant 99030445. 

\section*{Data Availability} 
The data underlying this article will be shared on reasonable request to the corresponding author.

\begin{appendix}

\section{List of references with evaluated mass function slopes for open clusters in common with this work}

\begin{table*}
    \centering
    \caption{PDMF of the OCs in the other works.}
    \label{table_comparison}
    \begin{tabular}{cccccccccc}  
        \hline     
        Cluster & Reference & Mass range        &  MF slope   \\
                &           & ($\text{M}_{\odot}$) &          \\
        \hline
        Hyades:    &                        &               &                  \\     
                   & \cite{roser2011}       & 0.9 -- 2      & $-2.7$           \\
                   &  inside tidal radius   & 0.25 - 0.9    & $-0.93$          \\
                   & \cite{goldman2013}*    & 0.13 -- 1.2   & $-1.15\pm 0.06$  \\
                   &                        & 0.80 -- 3.2   & $-3.53\pm 0.30$  \\
   Coma Berenices: &                        &               &                  \\
                   & \cite{kraus2007}       & 0.12 -- 1     & $-0.6\pm 0.3$    \\ 
                   & \cite{tang2018}        & 0.3 -- 2.3    & $-0.49\pm 0.03$  \\ 
                   & \cite{tang2019}        & 0.25 -- 2.51  & $-0.79\pm 0.16$  \\   
        Pleiades:  &                        &               &                  \\  
                   & \cite{bouvier1998}     & 0.04 -- 0.4   & $-0.6\pm 0.15$   \\
                   & \cite{sanner2001}*     & 1.0 -- 4.1    & $-2.99\pm 0.39$  \\
                   & \cite{dobbie2002}      & 0.03 -- 0.6   & $-0.8$            \\
                   & \cite{tej2002}         & 0.055 -- 0.5  & $-0.5\pm 0.2$    \\
                   & \cite{moraux2003}      & 0.03 -- 0.48  & $-0.6\pm 0.11$   \\
                   &                        & 1.5 -- 10     & $-2.7$           \\
                   & \cite{bouy2015}*       & 0.20 -- 0.60  & $-1.23$          \\
                   &                        & 1.58 -- 4.79  & $-3.56$          \\
                   & \cite{olivares2018}    & 0.2 -- 0.56   & $-1.12\pm 0.08$  \\
                   & \cite{sollima2019}     & 0.25 -- 1     & $-1.6\pm 0.2$    \\
                   &                        & 1 -- 1.8      & $-3.4\pm 0.3$    \\
                   & \cite{niu2020}         & 0.57 -- 1.00  & $-1.97\pm 0.22$  \\
                   &                        & 1.00 -- 3.75  & $-3.09\pm 0.18$  \\ 
                   & \cite{roser2020}       & 0.3 -- 1      & $-0.77$          \\
                   &                        & 1 -- 2.5      & $-2.41$          \\
       IC 2391:    &                        &               &                  \\
                   & \cite{sanner2001}*     & 1.1 -- 8.1    & $-2.07\pm 0.53$  \\
                   & \cite{barrado2004}     & 0.072 -- 0.5  & $-0.96\pm 0.12$  \\
                   & \cite{parker2013}      & 0.13 -- 1.0   & $-1.7\pm 0.4$    \\
      $\alpha$-Per:&                        &               &                  \\
                   & \cite{sanner2001}*     & 1.1 -- 6.8    & $-2.57\pm 0.44$  \\
                   & \cite{deacon2004}      & 0.2 -- 1      & $-0.86^{+0.14}_{-0.19}$ \\
                   & \cite{sheikhi2016}:    &               &                  \\
                   & Without correction for binares & 0.10 -- 0.62  & $-0.5\pm 0.09$   \\
                   &                                & 0.62 -- 4.68  & $-2.32\pm 0.14$  \\ 
                   & Corrected for binaries         & 0.10 -- 0.62  & $-0.89\pm 0.11$  \\
                   &                                & 0.62 -- 4.68  & $-2.37\pm 0.09$  \\
        Praesepe:  &                        &               &                  \\
                   & \cite{hambly1995}      & 0.1 -- 0.5    & $-1.5$            \\
                   & \cite{williams1995}    & 0.08 -- 1.4   & $-1.34\pm 0.25$          \\
                   & \cite{adams2002}       & 0.1 -- 0.4    & $-0.08\pm 0.03$  \\
                   &                        & 0.4 -- 1.0    & $-1.63\pm 0.07$  \\
                   & \cite{kraus2007}       & 0.12 -- 1     & $-1.4\pm 0.2$    \\
                   & \cite{baker2010}:      &               &                  \\
                   & Z band, $t_\text{age}=500$ Myr & 0.125 -- 0.6  & $-1.11\pm 0.37$  \\
                   & Z band, $t_\text{age}=1$ Gyr   & 0.125 -- 0.6  & $-1.10\pm 0.37$  \\
                   & J band, $t_\text{age}=500$ Myr, 1 Gyr & 0.125 -- 0.6  & $-1.07$   \\
                   & K band, $t_\text{age}=500$ Myr, 1 Gyr & 0.125 -- 0.6  & $-1.09$   \\
                   & \cite{boudreault2010}  & 0.1 -- 0.6   & $-1.8\pm 0.1$    \\
                   & \cite{khalaj2013}:     &               &                  \\
                   & Without correction for binaries    & 0.15 -- 0.65  & $-0.85\pm 0.10$  \\
                   &                                    & 0.65 -- 2.20 & $-2.88\pm 0.22$ \\
                   & With correction for binaries       & 0.15 -- 0.65  & $-1.05\pm 0.05$  \\                       
                   &                                    & 0.65 -- 2.20  & $-2.80\pm 0.05$  \\
        NGC 2451A: &                        &               &                  \\
                   & \cite{sanner2001}*     & 1.3 -- 6.8    & $-1.69\pm 0.63$  \\          
        \hline  
    \end{tabular} 
\end{table*}  
\begin{table*}
    \centering
    \contcaption{}
    \begin{tabular}{cccccccccc} 
         \hline     
        Cluster & Reference & Mass range        &  MF slope   \\
                &           & ($\text{M}_{\odot}$) &             \\
        \hline
        Blanco 1:  &                         &               &                  \\ 
                   & \cite{sanner2001}*      & 1.1 -- 4.8    & $-3.27\pm 0.70$  \\                 
                   & \cite{moraux2007}       & 0.03 -- 0.60  & $-0.69\pm 0.15$  \\
                   & \cite{casewell2012}      & 0.03 -- 0.60  & $-0.93\pm 0.11$  \\
                   & \cite{zhang2020}        & 0.25 -- 2.51  & $-1.35\pm 0.20$  \\
        NGC 7092:  &                         &               &                  \\
                   & \cite{sanner2001}*      & 1.4 -- 3.4    & $-2.93\pm 1.24$  \\           
        NGC 2547:  &                         &               &                  \\
                   & \cite{naylor2002}*:     &               &                  \\
                   & Models of \cite{d'antona1997}           & 0.1 -- 0.8 & $-1.20\pm 0.25$ \\
                   &                                         & 1 -- 6     & $-2.8\pm 0.4$   \\
                   & Models of \cite{siess2000}              & 0.1 -- 0.8 & $-1.20\pm 0.18$ \\
                   &                                         & 1 -- 6     & $-3.0\pm 0.4$   \\
                   & \cite{jeffries2004}*:                   &            &                 \\
                   & Models of \cite{d'antona1997}           & 0.05 -- 0.4& $0.06\pm 0.20$  \\
                   & Models of \cite{baraffe2002}         & 0.05 -- 0.4& $0.23\pm 0.25$  \\                                
                   & Both above models                       &0.15 -- 0.65& $-1$     \\                
        NGC 2516:  &                         &               &                  \\
                   & \cite{jeffries2001}*:  &               &                  \\
                   & Solar metallicity models of \cite{siess2000}                & 0.3 -- 0.7 & $-0.25\pm 0.20$ \\
                   &                                                             & 0.7 -- 3.0 & $-2.47\pm 0.11$ \\
                   & Half-solar metallicity models of \cite{siess2000}           & 0.3 -- 0.7 & $-0.51\pm 0.13$ \\
                   &                                                             & 0.7 -- 3.0 & $-2.67\pm 0.11$ \\
                   & Solar metallicity models of Models of \cite{d'antona1997}   & 0.3 -- 0.7 & $0.00\pm 0.13$  \\
                   &                                                             & 0.7 -- 3.0 & $-2.58\pm 0.10$ \\
                   & \cite{sung2002}*:                                           &            &                 \\
                   & Without corretion for binaries                              & 0.8 -- 5   & $-2.4\pm 0.3$   \\
                   & Corrected for binaries                                      & 0.8 -- 5   & $-3.0\pm 0.6$   \\
                   & \cite{bonatto2005}*    & 0.39 -- 0.90  & $-1.42\pm 0.22$  \\ 
                   &                        & 0.90 -- 4.17  & $-2.41\pm 0.11$  \\              
        \hline  
    \end{tabular} 
      \begin{tablenotes}
      \item \textbf{Notes.} The references marked by (*) have reported the slope of linear fit ($\Gamma$) on MFs in the logarithmic representation using $\text{d}N(\log m)/\text{d}\log m \propto m^{-\Gamma}$. We converted their MF slopes to those in our notation using $\alpha=-(\Gamma+1)$.   
      \end{tablenotes}
\end{table*}  
\end{appendix}
\bsp \label{lastpage} \end{document}